\documentclass[%
 reprint,
 amsmath,amssymb,
 aps,
]{revtex4-1}

\usepackage{graphicx}
\usepackage{dcolumn}
\usepackage{bm}
\usepackage{booktabs}
\usepackage{amsmath, amssymb}
\usepackage{type1cm}

\DeclareMathOperator*{\argmax}{arg\,max}

\begin{document}

\preprint{APS/123-QED}

\title{Effect of Burstiness on the Air Transportation System}

\author{Hidetaka Ito$^{1,}$}
\email{ito@jamology.rcast.u-tokyo.ac.jp}
\thanks{}%
\author{Katsuhiro Nishinari$^{2}$}%
\affiliation{%
$^{1}$ Department of Aeronautics and Astronautics, School of Engineering, The University of Tokyo,\\
7-3-1 Hongo, Bunkyo-ku, Tokyo 113-8656, Japan \\
$^{2}$ Research Center for Advanced Science and Technology, The University of Tokyo,\\
4-6-1 Komaba, Meguro-ku, Tokyo 153-8904, Japan
}%

\date{\today}

\begin{abstract}
The effect of burstiness in complex networks has received considerable attention. In particular, its effect on temporal distance and delays in the air transportation system is significant owing to  their huge impact on our society. Therefore, in this paper, we propose two indexes of temporal distance based on passengers' behavior and analyze the effect. As a result, we find that burstiness shortens the temporal distance while delays are increased. Moreover, we discover that the positive effect of burstiness is lost when flight schedules get overcrowded.

\end{abstract}

\maketitle


\section{INTRODUCTION}
\label{sec:1}
Considerable attention has been paid to properties of complex networks \cite{RevModPhys.74.47, RevModPhys.81.591, arenas2008synchronization, barthelemy2011spatial, kivela2014multilayer, pastor2015epidemic}. Physicists have recently discovered properties universally seen in artificial, human, and natural systems \cite{newman2003structure, boccaletti2006complex}. In particular, the temporal behavior of such systems is one of the most important issues in physics \cite{holme2012temporal}. Burstiness---a concentration of events---is a major property of temporal networks \cite{eckmann2004entropy}. It can be represented by the probability distribution function of the inter-event times following a power-law function with an exponential cutoff, $p(\tau) = e^{-\tau/\tau_0}\tau^{-\alpha}$, where $\alpha$ is the exponent of the power law \cite{barabasi2005origin, karsai2012universal}. Various phenomena such as human activities \cite{vazquez2006modeling, goh2008burstiness, cattuto2010dynamics} and natural phenomena \cite{kemuriyama2010power, saichev2006universal} have this property. Furthermore, the effect that burstiness has on a system has become of recent interest. Using empirical telephone call and e-mail data, Karsai {\it et al.} argue that burstiness slows the spread of epidemics \cite{karsai2011small}. Although the effect of burstiness has been empirically and analytically studied \cite{iribarren2009impact, vazquez2007impact2, jo2014analytically, gavalda2014impact, horvath2014spreading, PhysRevLett.114.108701}, much remains unknown so far.

\begin{figure}
\begin{center}
\includegraphics[width=6cm]{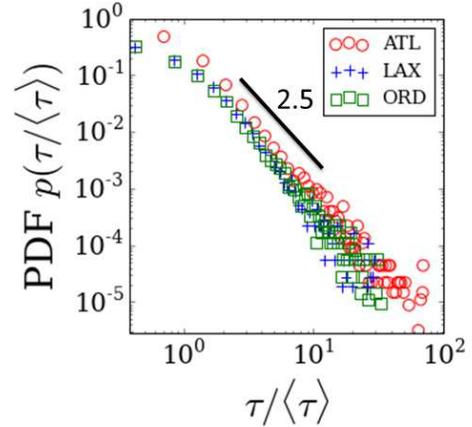}
\caption{(Color online) PDFs of the inter-arrival times at three major U.S. airports. The distributions follow power laws with an exponent of $\alpha = 2.5$ and a cutoff.}
\label{fig:18}
\end{center}
\end{figure}

U.S. flight schedules exhibit burstiness. Figure~\ref{fig:18} shows the probability density functions (PDFs) of inter-arrival times at three major U.S. airports. The inter-arrival time is the interval between two consecutive arrivals at the airport. If the distribution of the scheduled arrivalf times is smoothed, the PDFs of inter-arrival times follow exponential distributions. However, the PDFs follow the power-law distribution
\begin{equation}
p(\tau) \sim e^{-\tau/\tau_0} \tau^{-2.5},
\end{equation} 
where $\tau_0$ denotes the cutoff value. This result indicates that the U.S. air transportation system has burstiness \cite{Ito2015universal}. This property is universally seen for major airports in the United States. Likewise, the PDFs of inter-departure times also follow power laws with an exponent of $\alpha = 2.5$. See our previous work \cite{Ito2015universal} for further discussion.

\begin{figure}
\begin{center}
\includegraphics[width=7cm]{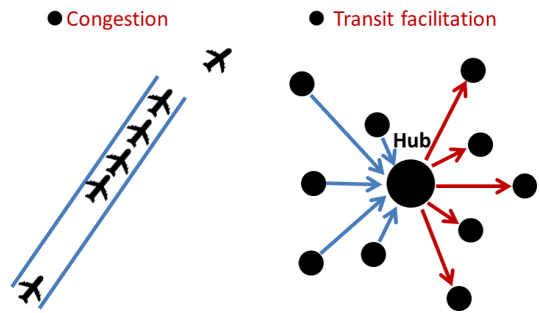}
\caption{(Color online) Schematic view of two major effects of burstiness on an air transportation system: congestion (a negative effect) and transit facilitation (a positive effect). The balance between the two effects determine whether burstiness is an effective measure for better air traffic.}
\label{fig:1}
\end{center}
\end{figure}

Because flight schedules are artificially constructed by airlines, burstiness in the air transportation system can be eliminated if burstiness is found to have an undue influence on the system. Then, the question arises as to whether flight schedules should have burstiness. Burstiness has two types of effects on the air transportation system, as shown in Fig.~\ref{fig:1}: congestion and transit facilitation. On one hand, burstiness contributes to severe congestion \cite{peterson1995models, ater2012internalization}. Burstiness clusters aircraft departing and arriving) at airports. Nonetheless, airport departure  and arrival capacities are limited \cite{gilbo1993airport}. Thus, a cluster of aircraft leads to more congested traffic. On the other hand, airlines adopt flight schedules with burstiness because burstiness facilitates plane connections \cite{bootsma1997airline, alderighi2007assessment, burghouwt2005temporal}. Since passengers travel to destinations via hub airports, plane connections play a major role in the air transportation system (see Appendix A for further details of the ratio of passengers' flight plans with plane connections). Thus, facilitating the transit of passengers at hub airports can increase flexibility of their travel plans and reduce their travel times. To provide passengers with many flight choices, numerous arriving and departing aircraft are arranged to be clustered. This narrows the distribution of passengers' connection times, which facilitates transit \cite{dennis1999competition}. The first effect of burstiness (congestion) is disadvantageous to the air transport system, but the second (transit facilitation) is advantageous. Therefore, which effect dominates is a controversial issue that requires study to determine whether burstiness is necessary for the air transportation system. Nonetheless, the effect of burstiness on the air transportation system is not well understood.

Meanwhile, analysis of complex systems and network theory have been applied to studies of various systems \cite{costa2011analyzing}. A main subject of the applications is traffic flows such as flows of vehicles, pedestrians, and aircraft \cite{chowdhury2000statistical, helbing2001traffic, bryan1999hub, guimera2005worldwide, karamouzas2014universal, li2015percolation}. Among them, the air transportation system is significant in today's global society. The ever-increasing number of flights has the system on the verge of dysfunction, with the number of flights having doubled every 15 years and being expected to double in the next 15 years \cite{Airbus}. Building new facilities to expand airports is a traditional way to resolve the problem. However, construction takes a long time and costs are substantial. Thus, a new measurement method to combat issues caused by the increasing number of flights without new construction of facilities is urgently needed.

Flights are delayed for various reasons. As the number of flights increases, routes and airports get congested \cite{gwiggner2014data, lacasa2009jamming, ezaki2015taming}. Other causes include weather, airlinesf troubles, airport operation, security, and late-arriving aircraft \cite{BTS, ahmadbeygi2008analysis, ahmadbeygi2010decreasing}. Thus, planning optimal flight schedules is a crucial issue \cite{barnhart2012demand}. Passengers' travel times and delays taking into consideration of passengers' plane connections are typical factors of optimality \cite{barnhart2014modeling}. The minimum travel time of a route when passengers start traveling is defined as the temporal distance in network science \cite{holme2015modern}. Pan {\it et al.} proposed an algorithm to calculate the temporal distance \cite{pan2011path}. Nonetheless, although delays, missed connections, and flight cancellation have tremendous impact on the temporal distance, which is the sum of the scheduled temporal distance and delays, Pan's algorithm does not take them into consideration. Therefore, it is necessary to extend the algorithm to consider them.

Therefore, in this paper, we discuss the effect of burstiness with an extended temporal graph analysis method. We find that the burstiness reduces the actual temporal distance if the starting times of traveling passengers are randomly distributed (as supposed for business travelers). However, burstiness only slightly decreases the temporal distance if passengers adopt a strategy to minimize the time they are forced to stay at airports and be on board (as supposed for tourists). Moreover, we simulate the case in which the airports are overcrowded because of a large number of flights. When the number of flights increases, the temporal distance is often larger in the case of bursty flight schedules than for nonbursty ones.

The remainder of this paper is organized as follows. In Sec.~\ref{sec:2}, we discuss the analysis and data-processing methods. In Sec.~\ref{sec:3}, we propose two indexes of the temporal distance based on passenger characteristics. In Sec.~\ref{sec:4}, we compare the indexes of temporal distances of the original and regularized (without burstiness) flight schedules. In Sec.~\ref{sec:5}, we discuss the case in which the number of flights increases. Section~\ref{sec:6} provides the conclusion. 

\section{METHODS}
\label{sec:2}

\subsection{Temporal distance}

We adopt an algorithm to calculate temporal distances at an arbitrary travel starting time from origins to destinations as proposed by Pan {\it et al.} \cite{pan2011path} (see Appendix B for details) while we implement some modifications for this paper. We assume that the passengers secure at least 45 min for making a plane connection. This indicates that a passenger can decide to use the flight as a connection flight if the connection flight's departure time is $\geq$45 min after the arrival time of the flight on which the passenger reaches the airport. (See Appendix C for a discussion on variations of secure times.) The temporal distances are given by the difference between passengers' arrival and departure times. In this paper, two types of temporal distances can be calculated: scheduled and actual ones. Schedules temporal distances are calculated by using scheduled arrival and departure times, but actual temporal distances are calculated by using actual arrival time and scheduled departure time in this paper. This is because passengers arrive at the origin airport independent of the flight's actual departure time. As a result of the calculation, we obtain the relationship between the time and temporal distance. In addition, missed connections and flight cancellations have to be considered in the calculation of actual temporal distances, which is discussed later.

\begin{figure}
\begin{center}
\includegraphics[width=7cm]{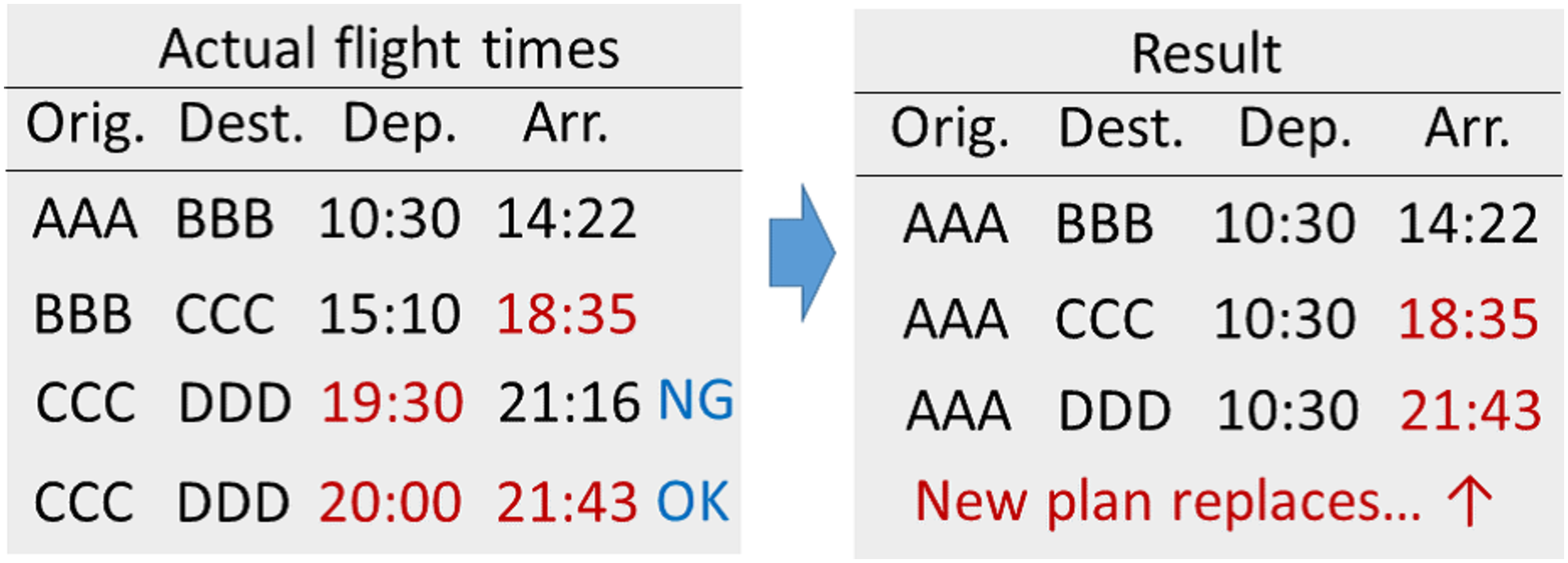}
\caption{(Color online) Schematic representation of the rules for handling missed connections. In the upper left panel, the actual flight schedules of three flights are shown. In the upper right panel, the flight results of passengers' initial plans for reaching BBB, CCC, and DDD from AAA with these three flights are shown. In the lower left panel, the actual flight schedules of candidate substitute flights for passengers missing their flights are shown. In the lower right panel, the results of the search for substitute flights are shown.}
\label{fig:20}
\end{center}
\end{figure}

\subsection{Delays}

In this section, we discuss the method used to calculate the delays. Passengers make their initial travel plans based on the departure and arrival times of the scheduled flight data. Delays occur when the actual arrival times of the flights are later than the scheduled ones. The arrival times of the passengers are often delayed because of missed connections and flight cancellations. Thus, plans with connections tend to be heavily delayed. (See Appendix D for a discussion of the relationship between the distribution of delays and the number of connections.) 

\subsubsection{Missed connections}

Let us now discuss the way to handle missed connections. We assume that passengers need 20 min to make a plane connection. Thus, if $<$ 20min is left for the plane connections at the time when passengers arrive at the airport, they miss their connections. If passengers miss connections, they search for substitute flights to reach their destinations as soon as possible. In the case of missed connections, we assume that it takes 60 min as a penalty. This penalty time corresponds to a time it takes to search for a new flight since missed connections are often sudden incidents. (See Appendix E for a discussion on the difference of the temporal distance in eliminating the penalty time.)

We now discuss a sample case for better understanding of the rule. We assume a situation in which four airports named AAA, BBB, CCC, and DDD are operated and passengers plan to reach their destinations (BBB, CCC, and DDD) from AAA. As shown in the upper left panel of Fig~\ref{fig:20}, AAA, BBB, and CCC have direct flights to BBB, CCC, and DDD, respectively. Thus, a passenger whose destination is DDD takes these three flights (AAA to BBB, BBB to CCC, and CCC to DDD) to reach the destination. Then, we assume that the flight from BBB to CCC is delayed and that the actual arrival time of the flight is 18:35. Since the departure time of the flight from CCC to DDD is 18:50, only 15 min is left for a plane connection. This indicates that the plan of the passenger from AAA to DDD is not feasible (NF), as shown in the upper right panel of Fig~\ref{fig:20}. Then, the passenger searches for a flight on which he or she can reach the destination as soon as possible. Two candidates of substitute flights are shown in the lower left panel of Fig~\ref{fig:20}. Since it takes 60 min to search for a new flight, the passenger cannot take a flight departing at 19:30. The passenger has to search for a substitute flight departing at CCC later than 19:35. Therefore, the passenger who misses the scheduled flight can take the flight departing at 20:00. As a result, the new plan replaces the original plan and the actual arrival time of the plan from AAA to DDD is 21:43, as shown in the lower right panel of Fig~\ref{fig:20}.

\subsubsection{Flight cancellations}

\begin{figure}
\begin{center}
\includegraphics[width=7cm]{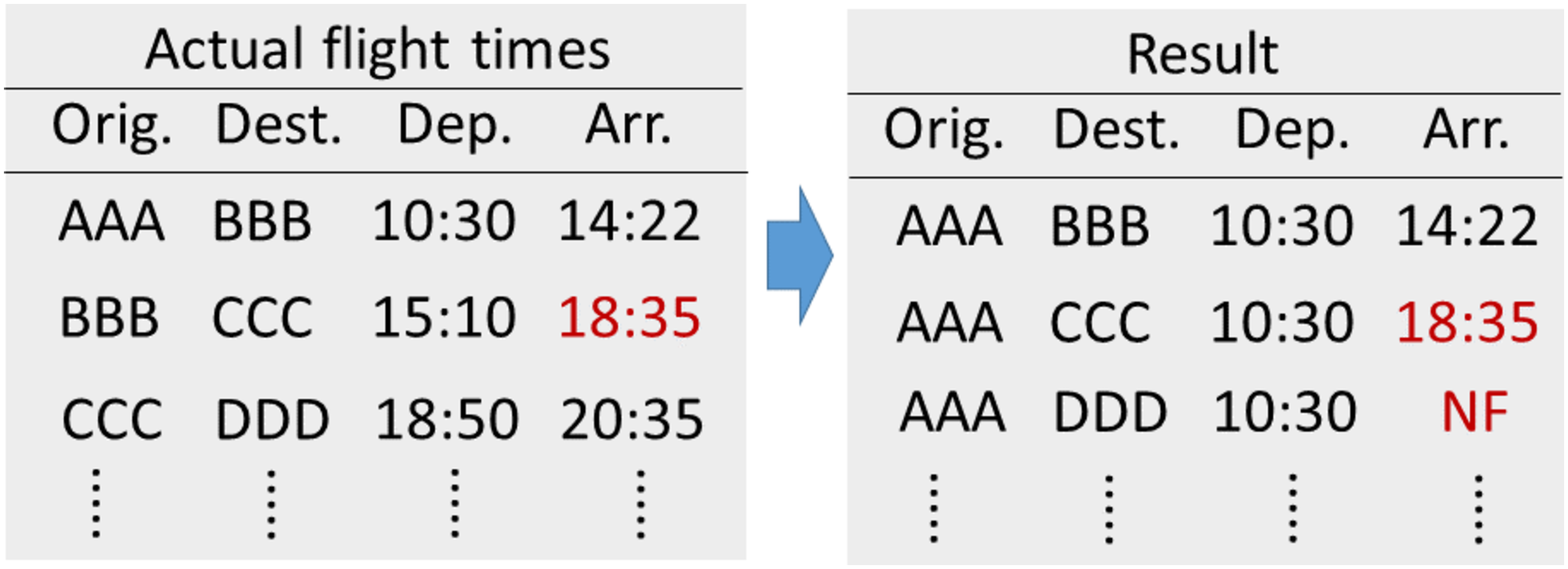}
\caption{(Color online) Schematic representation of the rules to handle flight cancellations. In the upper left panel, the actual flight schedules of three flights are shown. In the upper right panel, the flight results of passengers' initial plans for reaching BBB, CCC, and DDD from AAA with these three flights are shown. In the lower left panel, the actual flight schedules of candidate substitute flights for passengers whose initial flights are canceled are shown. In the lower right panel, the results of the search for substitute flights are shown. }
\label{fig:19}
\end{center}
\end{figure}

We now discuss the way to handle flight cancellations. If a flight is canceled, all passengers using the flight have to search for substitute flights to reach their destinations as soon as possible. In the case of cancellation, passengers can take flights immediately after the departure times of their initial plans' flights. This is because times between announcements of flight cancellations and departure times of these flights are long enough for passengers to search for new flight plans beforehand. Since data on the announcement time of flight cancellation are not available, we ignore the cases in which connection flights are suddenly canceled, in which case we should add penalty time in searching for new flight plans.

We discuss a sample case for better understanding of the rule. We assume the same situation previously discussed. As shown in the upper left panel of Fig~\ref{fig:19}, we assume that the flight from BBB to CCC is canceled. ``Cancel'' and ``NF'' indicates a canceled flight and a nonfeasible plan, respectively. Passengers who use this flight cannot reach their destinations unless they search for other flights. Two plans, from AAA to CCC and from AAA to DDD, are not feasible, as shown in upper right panel of Fig~\ref{fig:19}. Then, passengers search for flight plans that enable them to reach their destinations as soon as possible. In the case shown in the lower left panel of Fig.~\ref{fig:19}, the earliest flight from BBB to CCC is a flight departing and arriving at 16:31 and 20:28, respectively. Thus, the new plan replaces the scheduled plan. As a result, the actual arrival time of the plan from AAA to CCC at 10:30 is 20:28, as shown in the lower left panel of Fig.~\ref{fig:19}. The plan from AAA to DDD is also replaced following the same process.

\subsection{Regularization}
\label{sec:2-3}

In later sections, the temporal distance of the original flight schedule data is compared with that of the regularized flight schedule data. Here, we address how to regularize the flight schedule data. The flight schedules of 20 busy  airports with respect to the number of shortest temporal plans are regularized. First, the arrival flight schedule of the busiest airport is regularized. The flight schedule is divided into 2-hour flight schedules. The reason for using a 2-hour division is that, at the largest airport in the United States (ATL), the interval between the peaks in the number of  arrivals and departures is $\sim$2 hours. If the flight schedule is divided into too short of a time, burstiness cannot be eliminated. If the flight schedule is divided into too long of a time, the rough trend in the number of flights cannot be preserved. The rough trend indicates that the numbers of flights are large and small during day and night, respectively, which is different from burstiness. Having a large number of flights departing and arriving at night is not realistic since only a small number of passengers will take them. Therefore, the 2-hour division is appropriate for the regularization. Then, the numbers of arrival flights in the 2-hour flight schedules, $n$, are counted. Then, the regularized scheduled arrival time of the $i$th flights in a 2-hour flight schedule is given by 
\begin{equation}
T_i = T_{min} + \frac{i(T_{max} - T_{min})}{n},
\end{equation}
where $T_{min}$ and $T_{max}$ are the arrival times of the first and last flights in each 2-hours flight schedule, respectively. Setting maximum and minimum arrival times can prevent flights from being set to arrive when no flights are allowed to arrive owing to regulations. As an exception, to equally distribute the flight arrivals, the maximum time $T_{max}$ is set as the arrival time of the first flight in the flight schedule of the next 2 hours if it is divisible by 2 hours without a remainder. This regularization procedure preserves the order of arrivals. Because the resolution of the original flight schedule data is 1 min, the regularized arrival times are rounded off to the nearest whole number in minutes in this analysis. The departure times at the origins of their flights are shifted by the difference between the original and regularized arrival times. Then, the departure flight schedule of the busiest airport is regularized. The regularization procedure is the same as before. Then, regularization with the data from the next busiest airport is repeated until regularization of the flight schedule of all 20 airports is implemented.

\subsection{Airport capacity estimation and congestion}
\label{sec:2-4}

In later sections, we consider delays caused by congestion occurring around and in airports. Delays caused by congestion mainly depend on the number of arriving and departing aircraft, airport capacity (the number of runways), and weather. Delays caused by congestion are already included in the data in the case of the original schedule. However, delays caused by congestion are also changed if the flight schedules are regularized. We discuss the strategy to assess the delays in the case of the regularized schedules. 

First, we assess airport capacities. Airport capacities are defined as the times necessary for coping with all aircrafts. We assume that the airport is served using a queuing system having a constant service time. The amount of delay is given by the interval of time between the departure (arrival) and the beginning of service at the airport. Arrival and departure capacities are mutually correlated \cite{gilbo1993airport}. The total number of arrivals and departures are counted. For simplicity, we make the following assumptions: The times needed to manage a departure and arrival are the same. Every aircraft needs the same amount of time for departure (arrival). It is also assumed that each airport has a different capacity. The capacity of each airport is constant for 24 hours (from midnight to midnight). In addition, for simplicity, we do not consider each runway at an airport. We assume that each airport is assumed to have one runway with a capacity of all runways of the airport that handles all departures and arrivals. Next, we discuss the way to calculate airport capacities. Airport capacities are assessed based on the empirical data of the numbers of actual departures and arrivals. The empirical data of actual departures and arrivals are reliable since they show that the airport can actually manage departures and arrivals. We assume that the airport capacity $C$ is given by 
\begin{equation}
C = \min \left( \argmax_{t} N_{10}(t) , 0.5 \right),
\end{equation}
where $t$ and $N_{10}(t)$ are the time and the 10-min average of the total number of departures and arrivals, respectively. The runway capacity is not fully utilized owing to the small numbers of departures and arrivals if the airport capacity assessed with the empirical data is too small. Thus, the airport capacity is set to not less than 0.5 aircraft per minute even if the assessed airport capacity is small.

Second, we discuss the way to assess the total numbers of departures and arrivals at a given time. The total numbers of departures and arrivals are calculated based on the empirical data of the scheduled flight times. The actual flight schedule is not appropriate since the concentration of arrivals and departures is smoothed out by management of the traffic flow. However, airplanes seldom depart  or arrive at airports at their scheduled times. According to analysis of the empirical data \cite{mueller2002analysis}, the en-route delay time distribution can be modeled by a normal distribution with a mean of $2.46$ min and a standard deviation of $7.38$ min. Thus, with the scheduled departure (arrival) time of the $i$th aircraft, $t_{i,sch}$, and the normal distribution $\mathcal{N}(\mu_{delay}, \sigma^2)$, we assume that the departure (arrival) time of the $i$th aircraft, $t_{i,ass}$, is given by 
\begin{equation}
t_{i,ass} = t_{i,sch} + \mathcal{N}(\mu_{delay}, \sigma^2),
\end{equation}
where $\mu_{delay} = 2.46~{\rm min}$ and $\sigma = 7.38~{\rm min}$, respectively. Then, the departure (arrival) delay $d_{i}$ is given by 
\begin{equation}
d_{i} = \max (0, t_{i-1,ass} + 1/C - t_{i,ass}).
\end{equation}

Third, we discuss how to calculate the departure (arrival) time of the regularized flight schedules with the original flight schedule data. The original data include delays caused by congestion. Thus, using the method mentioned above, we subtract the delays calculated with the original data from the actual departure (arrival) time of the original data. In the case of departure time, the arrival time is also shifted by the delay of the departure times. In addition, departure (arrival) times are rounded off to the nearest whole number since the time resolution of the data is 1 min. Next, we perform regularization with this data. Then, we add the delays calculated using the actual departure (arrival) time of the regularized data. Similarly, the arrival time is also shifted by the delay of the departure times. In addition, departure (arrival) times are rounded off. The calculated departure (arrival) times are used as the actual departure (arrival) times of the regularized flight schedules.

\subsection{Data}

The data used in this analysis have been extracted from the Bureau of Transportation Statistics (BTS) \cite{BTS}. The Airline On-Time Performance Data and market data of the Airline Origin and Destination Survey recorded in 2014 have been analyzed in this paper. The Airline On-Time Performance Data include data from domestic flights reported by carriers with that each have at least a 1

\section{INDEXES OF TEMPORAL DISTANCE}
\label{sec:3}

\begin{figure}
\begin{center}
\includegraphics[width=8cm]{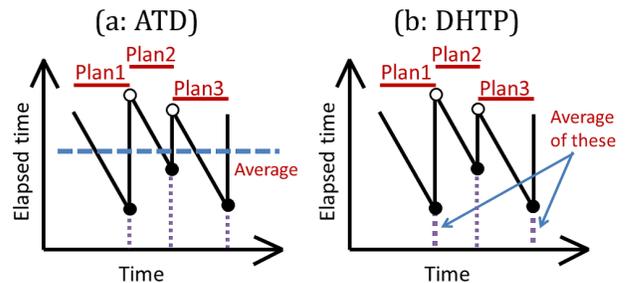}
\caption{(Color online) Schematic view of the definition of the indexes (a) ATD and (b) DHTP. The horizontal and vertical axes indicate the time and the temporal distance at that time, respectively. Black solid lines indicate the temporal distance. Red solid horizontal lines indicate the interval in which passengers use a plan. Purple dotted lines indicate the minimum temporal distance of plans. The ATD is given by the average of the temporal distance shown as a blue dashed line. The DHTP is given by the average of the minimum temporal distance of plans whose minimum temporal distances are less than or equal to the median.}
\label{fig:2}
\end{center}
\end{figure}

We propose two indexes for the temporal distance of the flight schedules based on passenger characteristics in this section. In the main sections, we discuss the average of temporal distance of the air transportation network as a whole. (See Appendix F for a discussion of the analysis of temporal distances at specific airports.)  Figure~\ref{fig:2} shows a schematic view of the definition of the indexes. The horizontal and vertical axes indicate the time at which passengers are able to depart an airport and the temporal distance, respectively. The temporal distance is the elapsed time for passengers to reach their destination. The slopes of the lines are $-1$ since passengers must arrive at airports before the departure times of flights.

The first index is the average temporal distance (ATD) proposed by Pan {\it et al.} \cite{pan2011path}. The ATD is defined as the average temporal distance at all time. We assume a periodic boundary condition, which indicates that the first flight departs again after departure of the last flight in the dataset. The ATD is given by
\begin{eqnarray}
{\rm ATD} &=& \frac{1}{T}\left[ t_1 \left( \frac{t_1}{2}+\delta t_1 \right) + (t_2 - t_1) \left( \frac{t_2 - t_1}{2}+\delta t_2 \right) \right. \nonumber \\
&+& \cdots + (t_n - t_{n-1}) \left( \frac{t_n - t_{n-1}}{2}+\delta t_n \right) \nonumber \\
&+& \left. (T-t_n) \left( \frac{T - t_n}{2}+ t_1 + \delta t_1 \right) \right],
\end{eqnarray}
where $T$, $t_i$, and $\delta t_i$ denote the total time of the data, the departure time, and the minimum temporal distance of the $i$th shortest plan, respectively \cite{pan2011path}. In the analysis used in this paper, the ATD indicates the average of all temporal distances of flight plans in the case in which the passengers are able to leave the airport at a randomly selected time with the limit of the number of passengers set to infinity. This index shows the temporal distance for a situation in which passengers want to reach their destination as soon as possible (e.g., as expected for business travelers). 

The second index is the average distance of half of the total plans (DHTP). The DHTP is given by 
\begin{equation}
{\rm DHTP} = 
\begin{cases}
\frac{\delta t_1 + \delta t_2 + \cdots + \delta t_{n/2}}{n/2} & (n\!: {\rm even}), \\
\frac{\delta t_1 + \delta t_2 + \cdots + \delta t_{n/2+1}}{n/2+1} & (n\!: {\rm odd}),
\end{cases}
\end{equation}
where $n$ denotes the number of the plans. The DHTP indicates the average temporal distance in the case in which passengers randomly choose one plan whose temporal distance is not larger than the median of all plans with the limit of the number of passengers set to infinity. The index shows the temporal distance for a situation in which passengers want to minimize the time they stay at the airport and are on board but they do not need to reach their destination as soon as possible (e.g., as expected for tourists). We discuss both ATD and DHTP in later sections. Later, we introduce temporal distance indexes to incorporate these two indexes (ATD and DHTP) as a whole.

\begin{figure}
\begin{center}
\includegraphics[width=8cm]{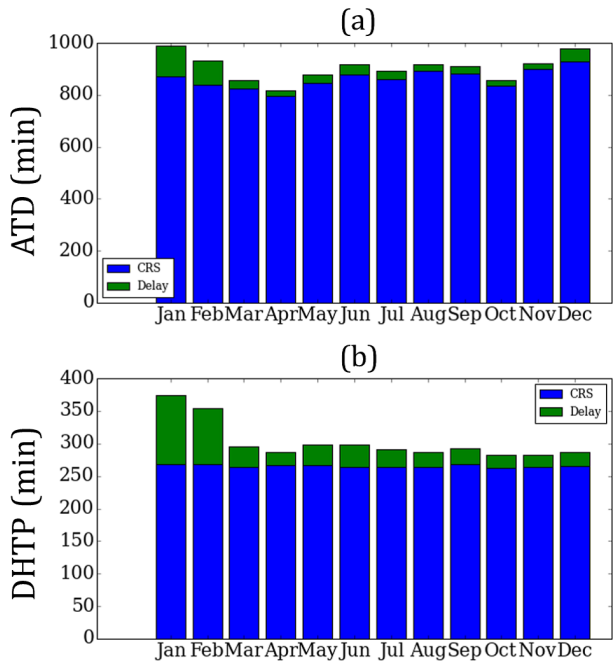}
\caption{(Color online) Monthly averages of the ATDs and DHTPs. Blue (lower) and green (upper) bars indicate the indexes of scheduled temporal distances and delays, respectively. The total heights of the bars indicates the indexes of actual temporal distances.}
\label{fig:16}
\end{center}
\end{figure}
 
Figure~\ref{fig:16} shows the monthly averages of the ATDs and DHTPs as stacked bar charts. These indexes are the weighted average of the temporal distance indexes (TDIs) of all passengers' flight routes, which indicates that popular routes are heavily weighted. Blue (lower) and green (upper) bars indicate the scheduled TDIs and delays, respectively. On one hand, ATDs vary by month, and the scheduled ATDs are minimum and maximum in April and December, respectively. On the other hand, DHTPs are independent of the month. This is because the minimum temporal distances of the plans are constant but the number and the intervals of the plans vary by month. Moreover, the delays caused in January and February are larger than those in other months. In January 2014, many flights are canceled, which affects the result. Additionally, the ratios of delays of ATDs and DHTPs are smaller and larger in May and December, respectively.

\section{QUANTIFICATION OF BURSTINESS EFFECT}
\label{sec:4}

\subsection{Temporal distance indexes}

\begin{figure}
\begin{center}
\includegraphics[width=8cm]{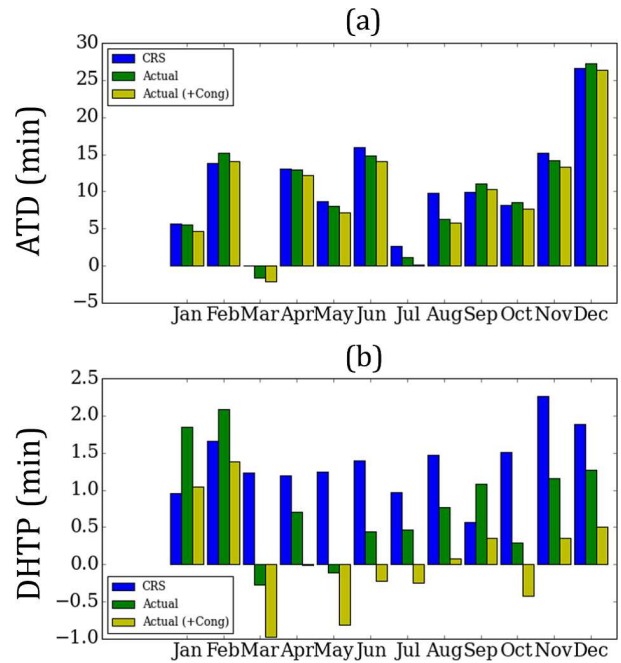}
\caption{(Color online) Difference in the (a) ATDs and (b) DHTPs between the original and regularized flight schedules. Positive values indicate that the TDIs of the original flight schedules are shorter than those of the regularized ones. ``CRS'' and ``Actual'' indicate the scheduled and actual flight schedules. ``Actual (+Cong) also indicates the actual flight schedules but the delays are altered by considering congestion caused by burstiness.}
\label{fig:11}
\end{center}
\end{figure}

In this section, we discuss the effect of burstiness on the temporal distances. We implement regularization of the flight schedules to simulate flight schedules without burstiness. Regularization is a process that makes flight schedules nonbursty as we discussed in Sec. \ref{sec:2-3}

Figure~\ref{fig:11} shows the differences in the (a) ATDs and (b) DHTPs between the original and regularized flight schedules. Positive values indicate that the TDIs of the original flight schedule are shorter. The blue (left) bar shows the TDIs of scheduled flight schedules. Green (center) and yellow (right) bars show the TDIs of the actual flight schedules, but only the yellow bar considers the congestion, which is represented by ``actual+congestion'' in this paper. Since bursty arrivals and departures are causes of congestion, as we discussed in the introduction, regularization reduces the delays caused by congestion, as we discussed in Sec. \ref{sec:2-4}

On one hand, the figure indicates that the ATDs of most months are reduced. The effect of congestion is overwhelmed by that of transit facilitation because of the well-designed flight schedule. (See Appendix E for a discussion on the elimination of the advantage of transit facilitation in the case of shuffled flight schedules.) The average reduction rates of the ATDs in the cases of scheduled, actual, and actual+congestion flight schedules are 1.23\%, 1.12\%, and 1.03\%, respectively. The reduction rate of the actual TDIs declines compared with the scheduled ones. This is because of a rise in the missed connection rates. In addition, the reduction rate of the actual+congestion TDIs declines compared with the actual ones. This is because of congestion caused by a concentration of departures and arrivals stemming from burstiness. However, the temporal distance is still reduced by $\sim$1.0\%, which significantly affects the drop of the time loss to passengers. Thus, burstiness has a positive effect in terms of the ATDs. 

On the other hand, the figure indicates that the DHTPs are not always reduced when delays and congestion are considered. The average reduction rates of the DHTPs in the cases of scheduled, actual, and actual+congestion flight schedules are 0.51\%, 0.26\%, and 0.02\%, respectively. The scheduled TDIs are always reduced, but the reduction rates of the actual+congestion DHTPs of 5 months out of 12 are negative. Among plans, plans with very short temporal distance have short connection times. Burstiness further shortens the connection time, which leads to missed connections. In addition, the DHTPs are heavily affected by the delays caused by congestion compared with the ATDs in terms of the ratio of delays to the TDIs. The delays caused by congestion make the DHTPs considerably larger. As a result, the actual DHTPs are not as reduced as we would expect in scheduling plans.

In summary, bursty departure and arrival behavior reduces the scheduled ATDs and DHTPs, but only ATDs are reduced when delays and congestion are considered. The reduction of the DHTPs obtained by burstiness is eliminated by the delays caused by missed connections and congestion.

\subsection{Delays}

\begin{figure}
\begin{center}
\includegraphics[width=8cm]{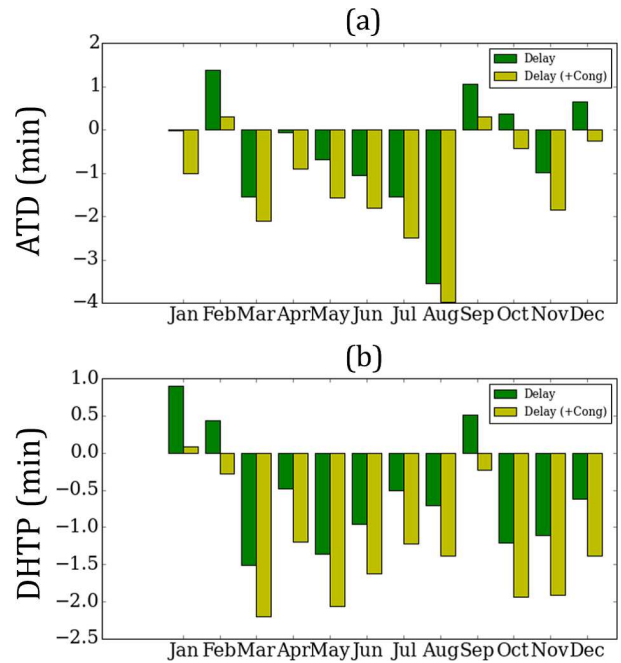}
\caption{(Color online) Difference in the delays of (a) ATDs and (b) DHTPs between the original and regularized flight schedules. Positive values indicate that the delays of the original flight schedules are shorter than those of the regularized ones. Delays in the case of ``Delay (+Cong)'' are altered by considering congestion caused by burstiness.}
\label{fig:13}
\end{center}
\end{figure}

We discuss the delays of the TDIs. Figure~\ref{fig:13} shows the differences in the delays of the (a) ATDs and (b) DHTPs between the original and regularized flight schedules. Positive values indicate that the delays of TDIs of the original flight schedule are smaller. Green (center) and yellow (right) bars show the TDIs without and with congestion, respectively. The latter is represented by ``actual+congestion'' in this paper.

The delays of  both the ATDs and DHTPs are increased. The increment rates of the ATDs' delays without and with congestion are 2.12\% and 4.53\%, respectively. The increment rates of the DHTPs' delays without and with congestion are 2.59\% and 5.26\%, respectively. The results show that the flight schedules with burstiness exhibit improvement in the TDIs at the expense of on-time performance.

\section{TOLERANCE TO INCREASING TRAFFIC }
\label{sec:5}

\begin{figure}
\begin{center}
\includegraphics[width=8.5cm]{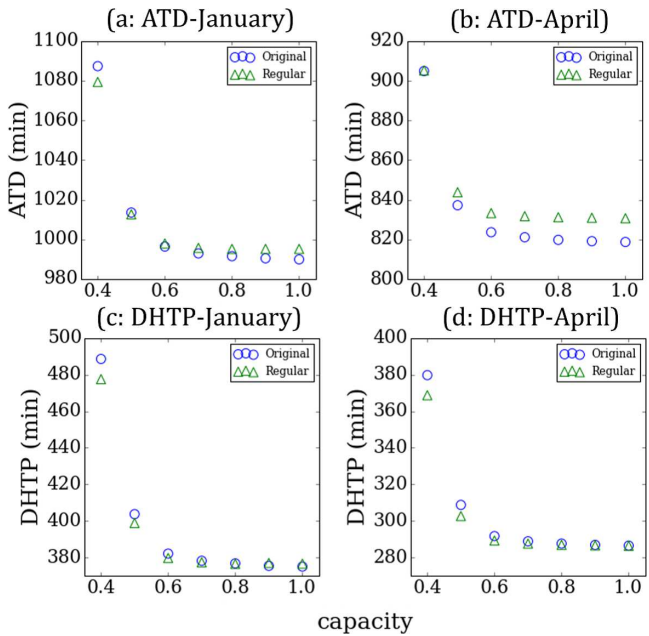}
\caption{(Color online)  Relationships between airport capacities and efficiency of the flight schedules in (a and b) January and (c and d) April. The vertical and horizontal axes are the TDIs and the ratio of the simulated capacities to the original ones, respectively.}
\label{fig:15}
\end{center}
\end{figure}

In this section, we discuss the tolerance of the original and regularized flight schedules to a rise in the amount of traffic. The ATDs and DHTPs in the cases of small airport capacities are investigated in this paper. The increase in the traffic amounts and the decreases in airport capacities are not the same. Nonetheless, the latter can simulate the former since their ratio is significant in discussing tolerance to congestion, which plays a main role in delays caused by changes in traffic volume. Moreover, adding new flights may alter the departure and arrival behavior in terms of origins and destinations if these new flights are not appropriately configured. Simulation with small airport capacities can avoid this risk. Thus, we study these cases from now  on. Figure~\ref{fig:15} shows the relationships between  airport capacities and TDIs in (a and b) January and (c and d) April. The vertical axes of Figs.~\ref{fig:15}(a) and (c) and Figs.~\ref{fig:15}(b) and (d) are the ATDs and DHTPs, respectively. The horizontal axes are the ratios of the simulated capacities to the original capacities; i.e., the simulated capacities $C'$ are given by $C' = C \times x$. Circles and triangles are original and regularized data, respectively.

The figures indicate that the original flight schedules are vulnerable to congestion caused by decreased capacity (which is equivalent to increased traffic). This property is commonly seen in the analysis of both ATDs and DHTPs. The lower the airport capacity is, the bigger is the difference in the delays caused by congestion between the original and regularized schedules. This result indicates that the U.S. air transportation system, which has bursty behavior, will experience severe congestion in the future because of increasing air traffic volume. Moreover, although the TDIs of the original schedules are smaller than the regularized ones in all four cases, those of the original schedules are larger than the regularized ones except the ATD in April when capacity is decreased. With regard to the ATD in April, the difference in the ATD between actual and regularized schedules is almost zero. Therefore, although flight schedules with burstiness function as a traffic flow enhancer now, they will not effectively function as such in the future. In addition, this future case has another serious problem: longer delays. This will lead to a huge inconvenience for passengers. Thus, future flight schedules should be devised without burstiness.

\section{CONCLUSION}
\label{sec:6}

In this paper, we proposed a method for calculating both the scheduled and actual temporal distance, considering delays, missed connections, and flight cancellations. Moreover, we discussed the efficacy of burstiness of the flight schedules in terms of temporal distance indexes (TDIs) by comparing the original and regularized flight schedules using the method. As TDIs, the average temporal distances (ATDs) and the average distances of half of the total shortest plans (DHTPs) are utilized. ATDs and DHTPs indicate the average of all passengers' flight plans for business travelers and tourists, respectively. As a result, we showed that the ATDs were shortened because burstiness facilitated plane connections and enabled more passengers to make connections with appropriate connection times. In addition, DHTPs were only slightly shortened because the decrease of the scheduled DHTPs was eliminated by the increase of the delays caused by missed connections and congestion. Therefore, in terms of the TDIs, burstiness is very effective for business passengers but not so efficacious for leisure passengers. Furthermore, we discussed the effect of a future traffic volume increase by simulation of small airport capacities. A small airport capacity corresponds to a high traffic volume. The result showed that the original scheduling is vulnerable to delays caused by congestion in the case of small airport capacity. In addition, although the TDIs of the original schedules were shorter than those of regularized schedules in all cases with normal capacity, the magnitudes of the TDIs were reversed as airport capacity decreases in three out of the four cases simulated in this work.

Burstiness is inherent in many systems and works as a system enhancer. This work has addressed the effect of burstiness on the air transportation system. In todayfs U.S. air transportation system, burstiness is advantageous to the system in terms of passengers' travel times. However, our work suggests that this effect would be lost as the number of future flights increases. The next generation of flight schedules should be designed to take this fact into account.

\section*{ACKNOWLEDGMENTS}
This work was supported by a Grant-in-Aid for Scientific Research (No. 25287026) from the Japan Society for the Promotion of Science. We wish to thank Daichi Yanagisawa and Takahiro Ezaki for their valuable comments on this manuscript.

\section*{APPENDIX A: DISTRIBUTION OF NUMBER OF FLIGHTS IN PLANS}

\begin{figure}
\begin{center}
\includegraphics[width=6cm]{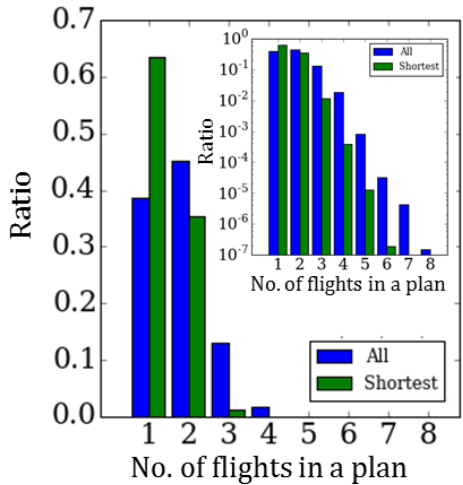}
\caption{(Color online) Ratio of the number of flights in plans on the basis of the number of passengers in January 2014. The inset shows the same data plotted on a log-log graph. Blue (left) and green (right) indicate the cases in which all passengers using a route are evenly distributed to all plans and plans with minimum connection times of the route, respectively.}
\label{fig:3}
\end{center}
\end{figure}

First, we discuss the number of flights in plans from an airport to another airport at an arbitrary time, which we obtain with the algorithm mentioned in Sec~\ref{sec:2}. Some plans consist of only one flight (i.e., a direct flight), but others have more than one flight. The ratio of the number of flights is important for the effect of burstiness on transit facilitation. Figure~\ref{fig:3} shows the relationship between the number of flights in plans and the ratio of flight plans in January 2014. The inset shows the same data on a log-log graph. We obtain the ratio by taking a weighted average based on the number of passengers utilizing the route. In paths from an origin to a destination, there are various plans with different numbers of flights. Blue (left) indicates the case in which all passengers using a route are evenly distributed to all plans of the route while green (right) bars show the cases in which all passengers using a route are evenly distributed to plans with minimum connection times among all plans of the route. Approximately 61\% and 37\% of passengers make plane connections based on the all plans and plans with the minimum number of flights, respectively. The actual ratio of plans with connections should be between these two cases for typical routes. A high ratio of plans with connection flights contributes to the enhancement of traffic by facilitating transit.

\begin{figure}
\begin{center}
\includegraphics[width=8.5cm]{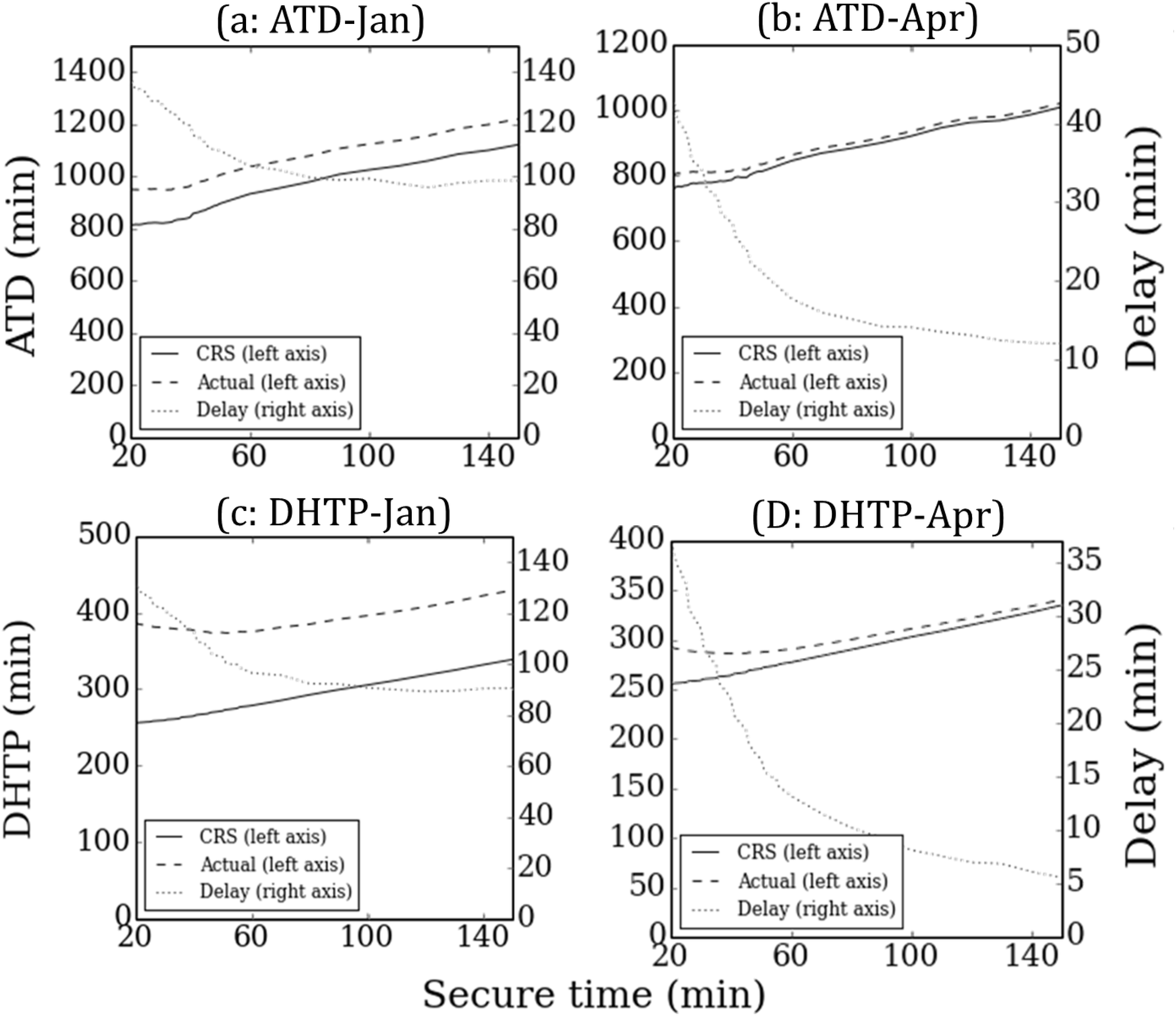}
\caption{(Color online) ATDs and DHTPs as a function of the secure times in January and April 2014, respectively. Solid, dashed, and dotted lines indicate the scheduled and actual TDIs and delays, respectively.}
\label{fig:10}
\end{center}
\end{figure}

\begin{figure}
\begin{center}
\includegraphics[width=8.5cm]{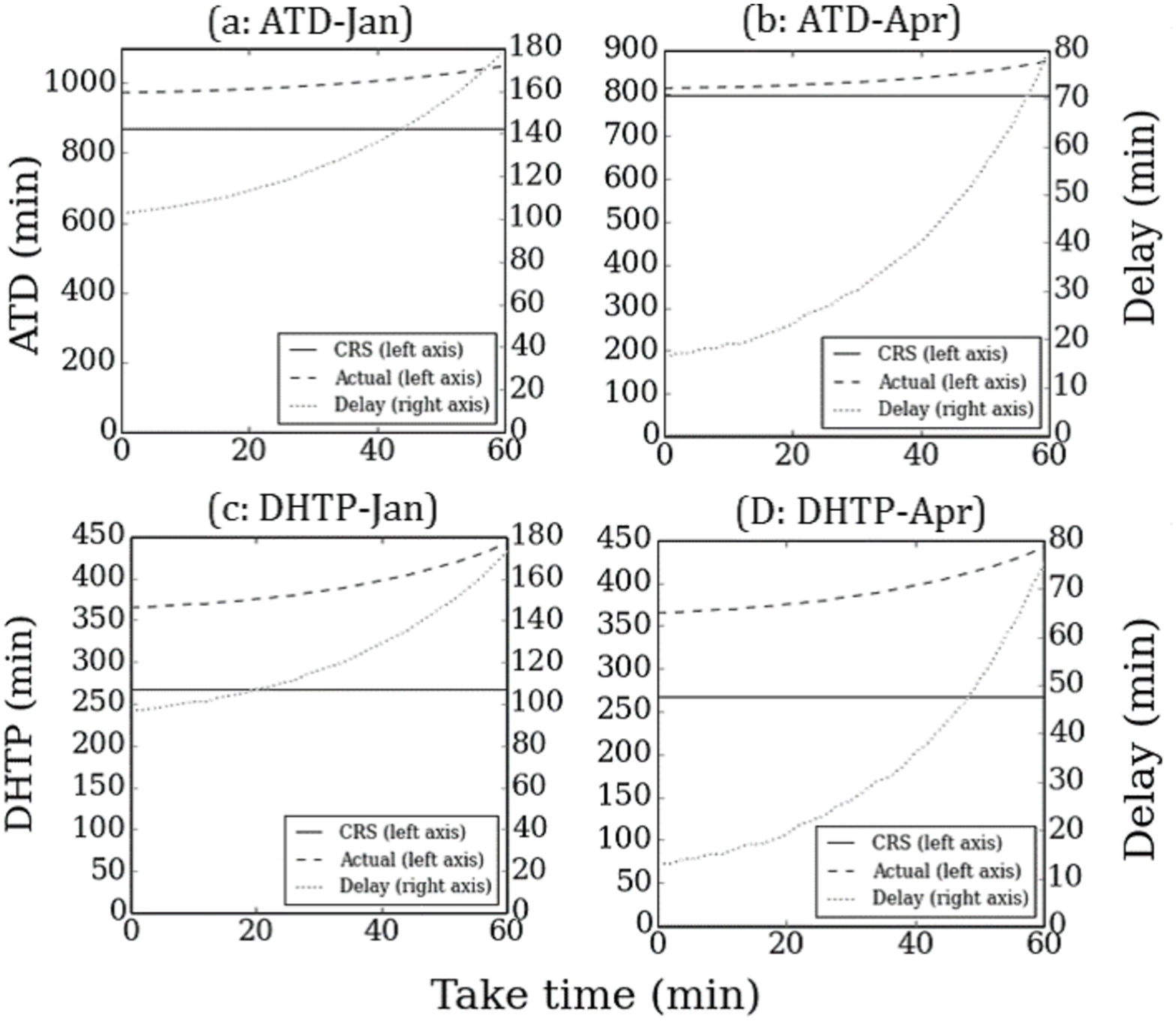}
\caption{(Color online) ATDs and DHTPs as a function of the take times in January and April 2014, respectively. Solid, dashed, and dotted lines indicate the scheduled and actual TDIs and delays, respectively.}
\label{fig:21}
\end{center}
\end{figure}

\begin{figure}
\begin{center}
\includegraphics[width=8.5cm]{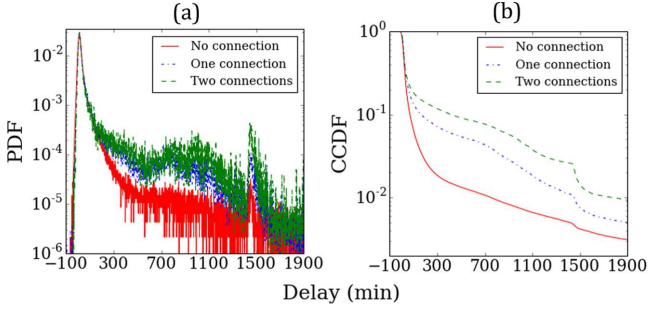}
\caption{(Color online) (a) Probability distribution functions (PDFs) and (b) complementary cumulative distribution functions (CCDFs) of delays.}
\label{fig:4}
\end{center}
\end{figure}

\begin{figure}
\begin{center}
\includegraphics[width=8cm]{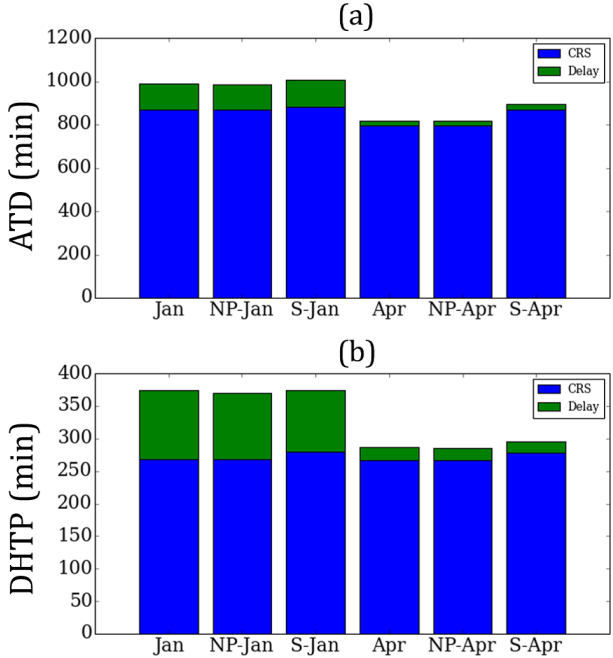}
\caption{(Color online) Monthly averages of (a) ATDs and (b) DHTPs with nonpenalty (NP) and shuffled (S) variations in January and April. The results of two variations are compared with the original data.}
\label{fig:17}
\end{center}
\end{figure}

\begin{figure}
\begin{center}
\includegraphics[width=8cm]{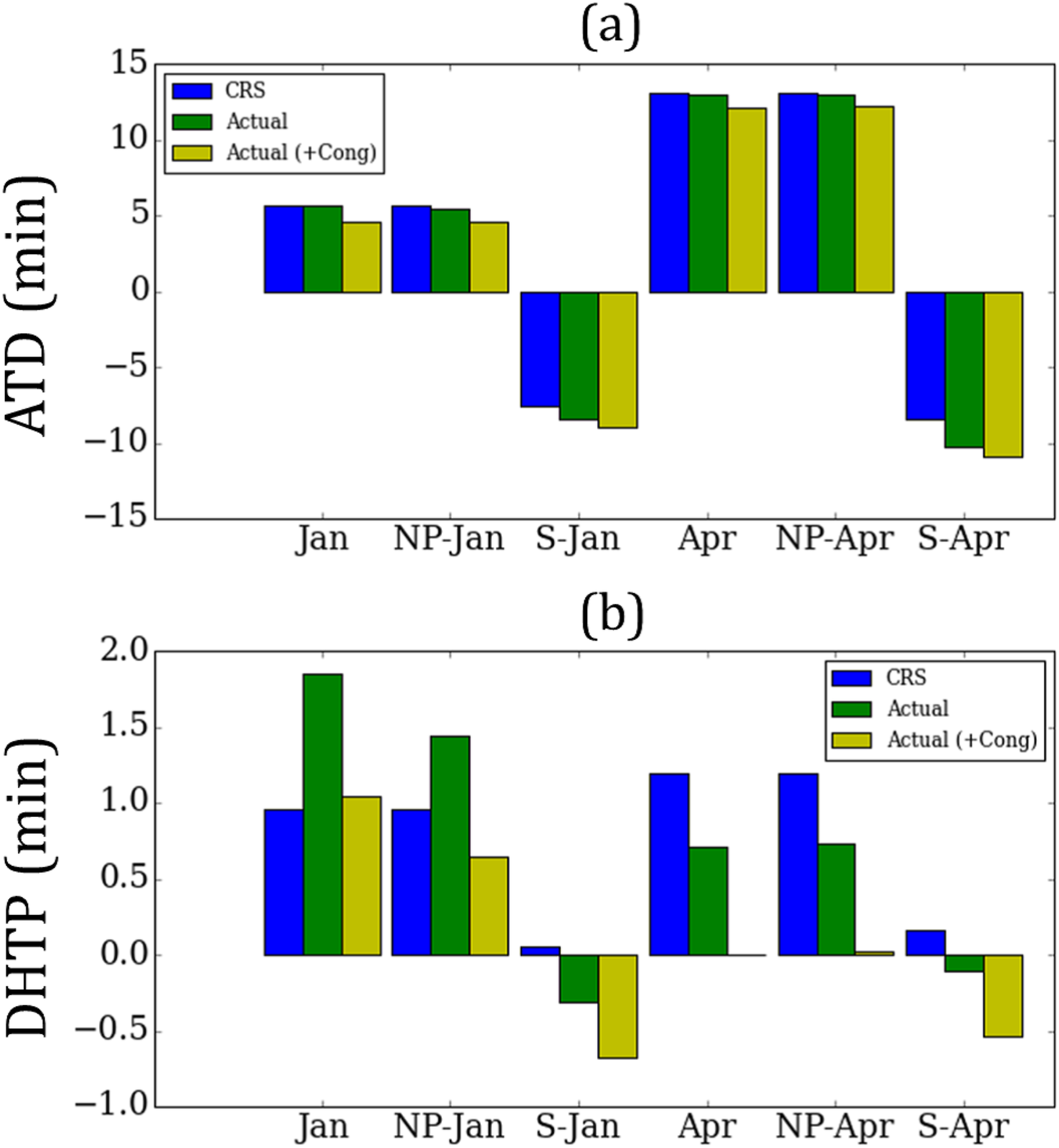}
\caption{(Color online) Differences in the (a) ATDs and (b) DHTPs between the original and regularized flight schedules with nonpenalty (NP) and shuffled (S) variations in January and April. The results of two variations are compared with the original data.}
\label{fig:12}
\end{center}
\end{figure}

\begin{figure}
\begin{center}
\includegraphics[width=8cm]{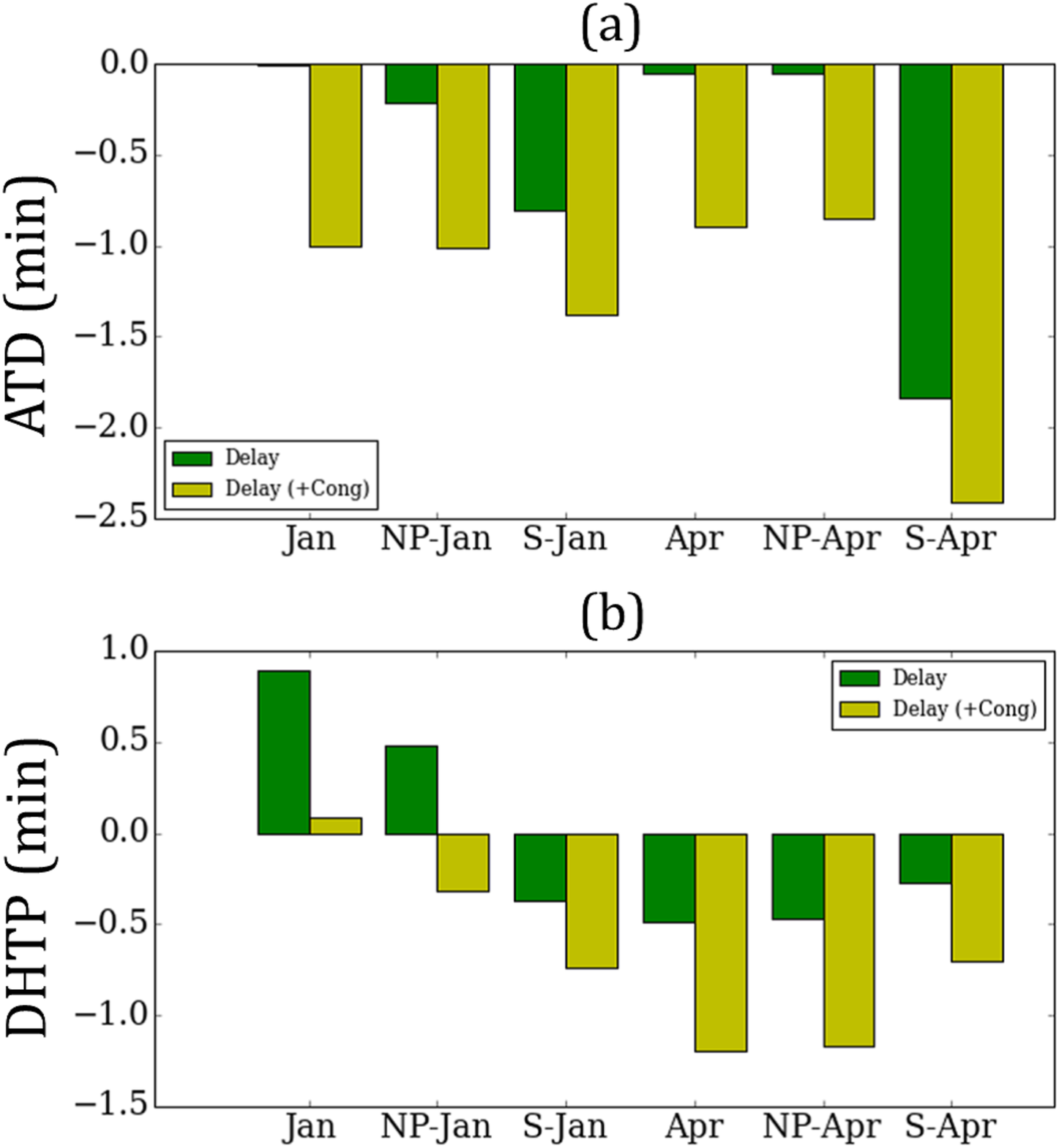}
\caption{(Color online) Differences in the delays of (a) ATDs and (b) DHTPs between the original and regularized flight schedules with nonpenalty (NP) and shuffled (S) variations in January and April. The results of two variations are compared with the original data.}
\label{fig:14}
\end{center}
\end{figure}

\begin{figure*}
\begin{center}
\includegraphics[width=18cm]{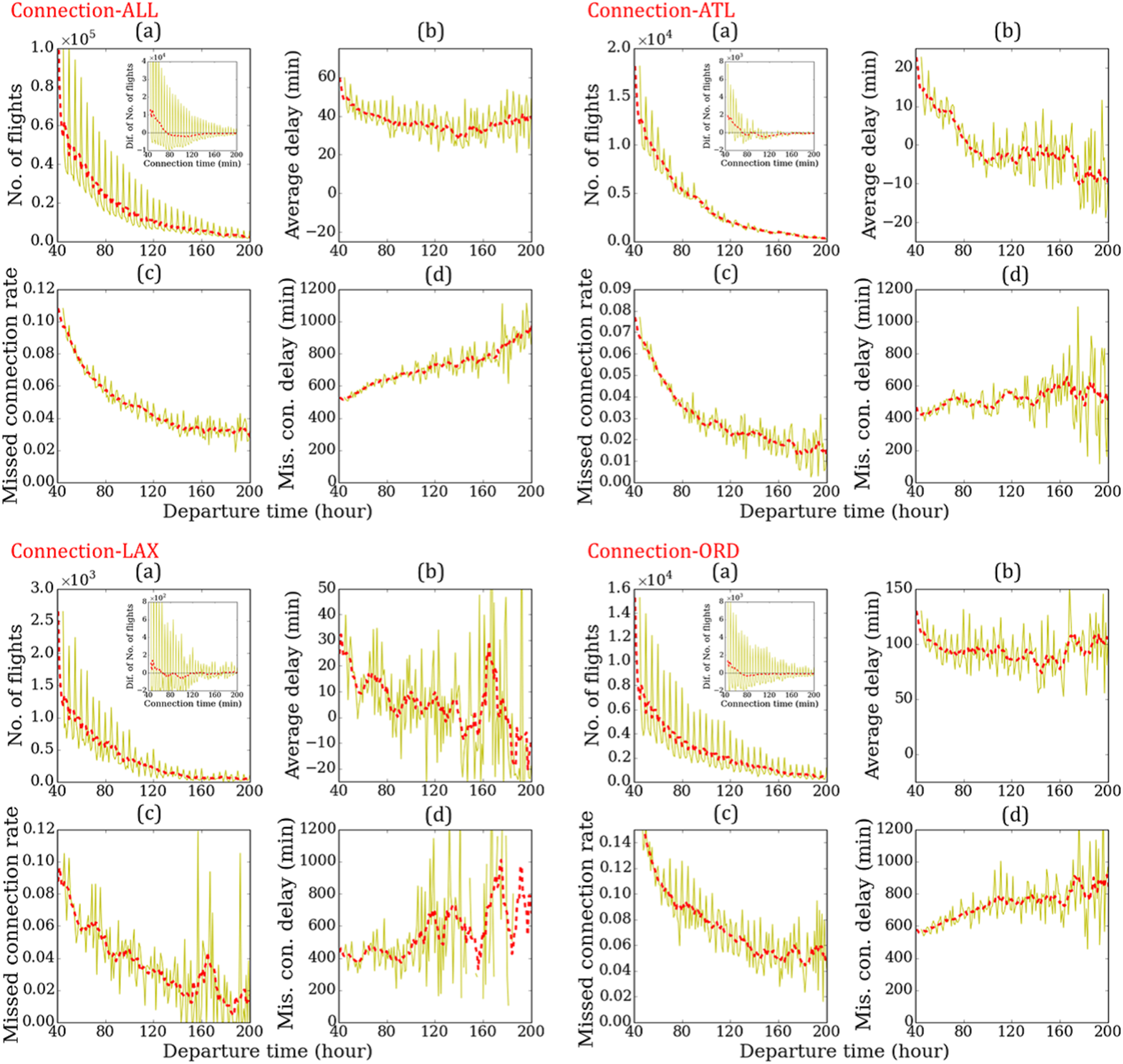}
\caption{(Color online) Relationships between the connection times and the (a) number of flights, (b) average delay, (c) missed connection rate, and (d) average delays in the missed connections at three main airports (ATL, LAX, and ORD) and aggregated data of all U.S. airports in January 2014. The inset of (a) shows the difference in the number of flights between the original and randomized data. Yellow solid and red dashed lines indicate the original data and a 9-min central moving average, respectively.}
\label{fig:6}
\end{center}
\end{figure*}

\begin{figure*}
\begin{center}
\includegraphics[width=18cm]{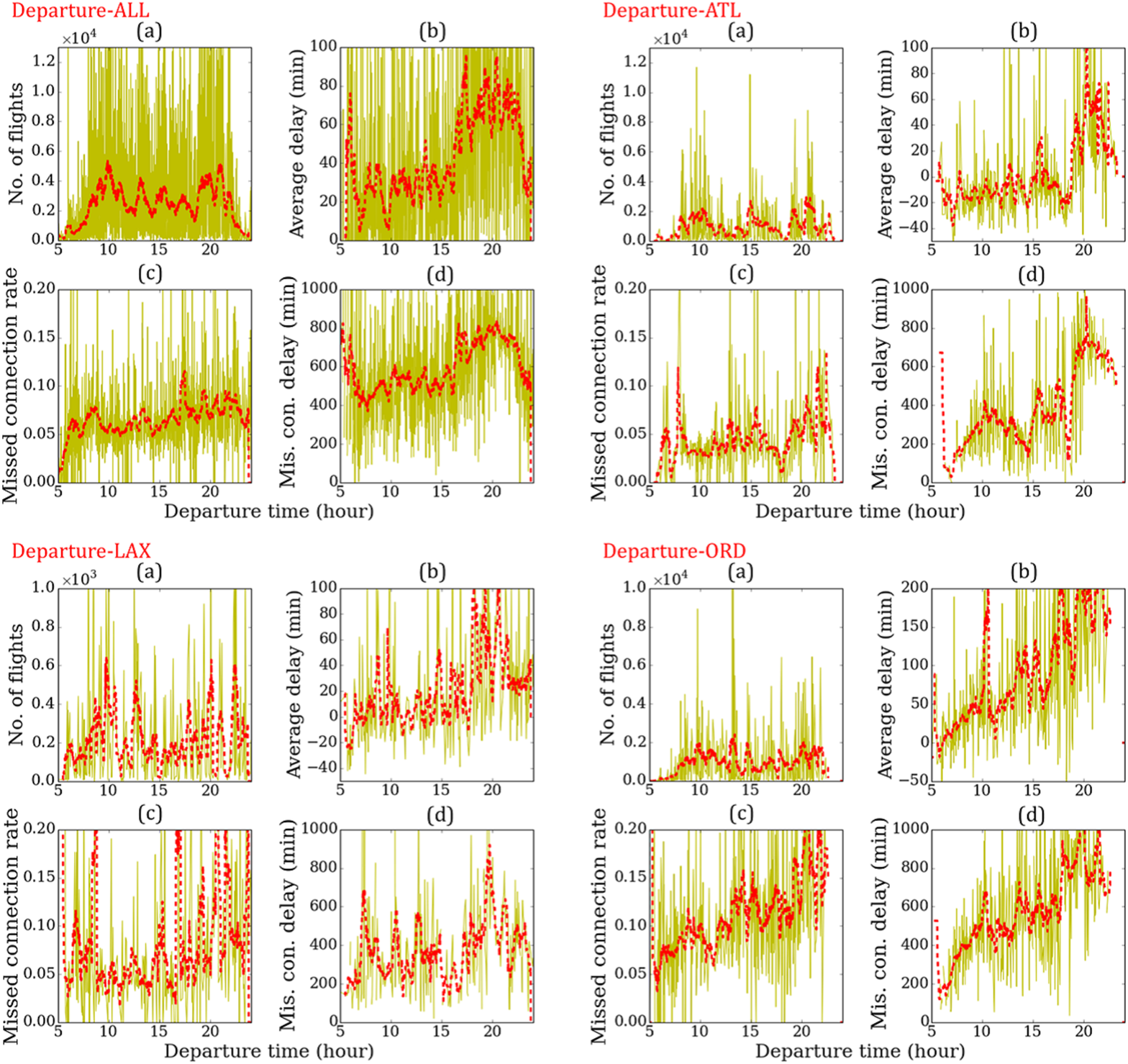}
\caption{(Color online) Relationships between the departure time and the (a) number of flights, (b) average delay, (c) missed connection rate, and (d) average delays in the case of the missed connections at three main airports and aggregated data of all U.S. airports in January 2014. Yellow solid and red dashed lines indicate the original data and a 29-min central moving average, respectively.}
\label{fig:5}
\end{center}
\end{figure*}

\begin{figure*}
\begin{center}
\includegraphics[width=18cm]{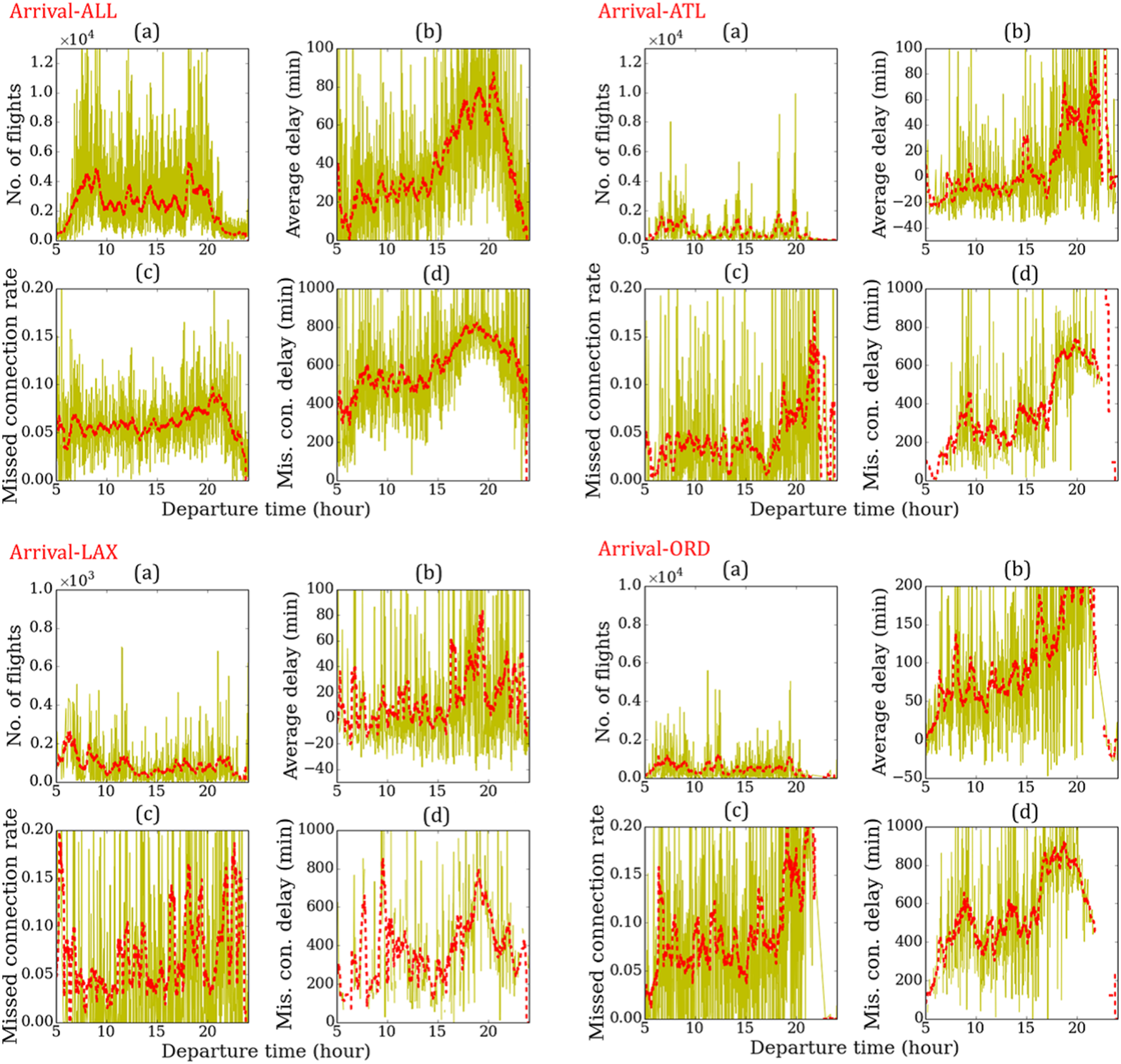}
\caption{(Color online) Relationships between the arrival time and the (a) number of flights, (b) average delay, (c) missed connection rate, and (d) average delays in the case of the missed connections at three main airports and aggregated data of all U.S. airports in January 2014. Yellow solid and red dashed lines indicate the original data and a 29-min central moving average, respectively.}
\label{fig:7}
\end{center}
\end{figure*}

\begin{figure*}
\begin{center}
\includegraphics[width=18cm]{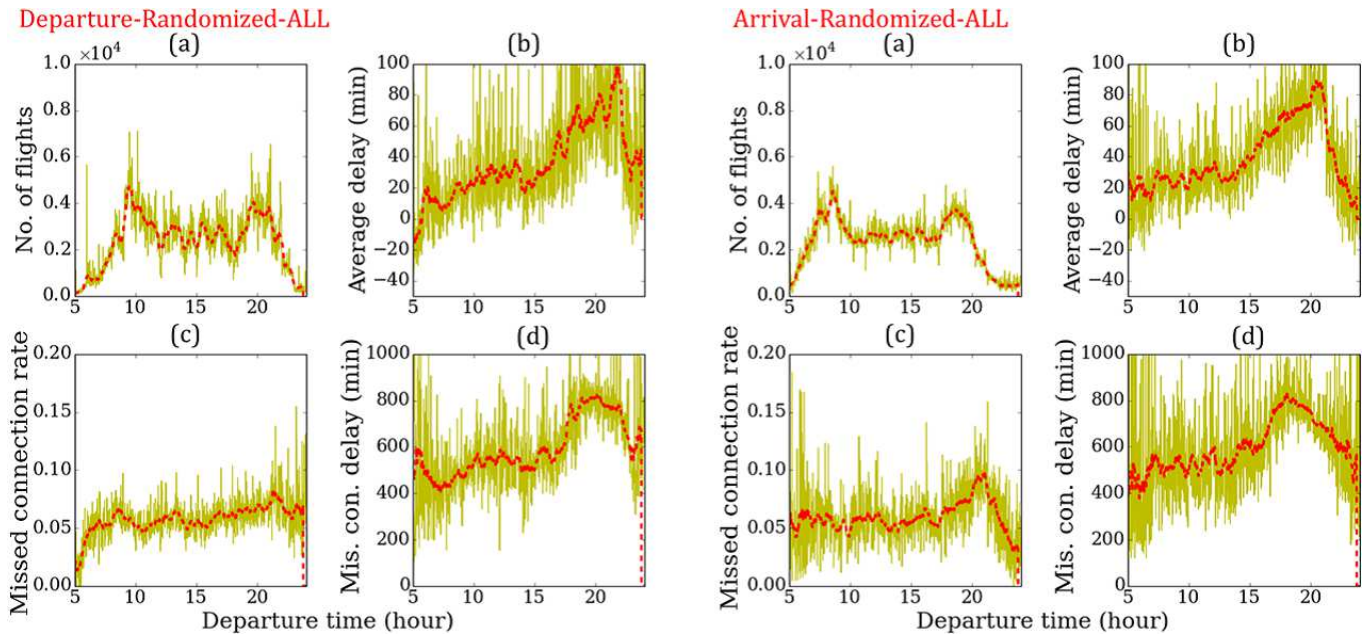}
\caption{(Color online) Relationships between the departure and arrival time and the (a) number of flights, (b) average delay, (c) missed connection rate, and (d) average delays in the case of the missed connections in aggregated data of all U.S. airports with the randomized data in January 2014. Yellow solid and red dashed lines indicate the original data and a 29-min central moving average, respectively.}
\label{fig:8}
\end{center}
\end{figure*}

\begin{figure*}
\begin{center}
\includegraphics[width=18cm]{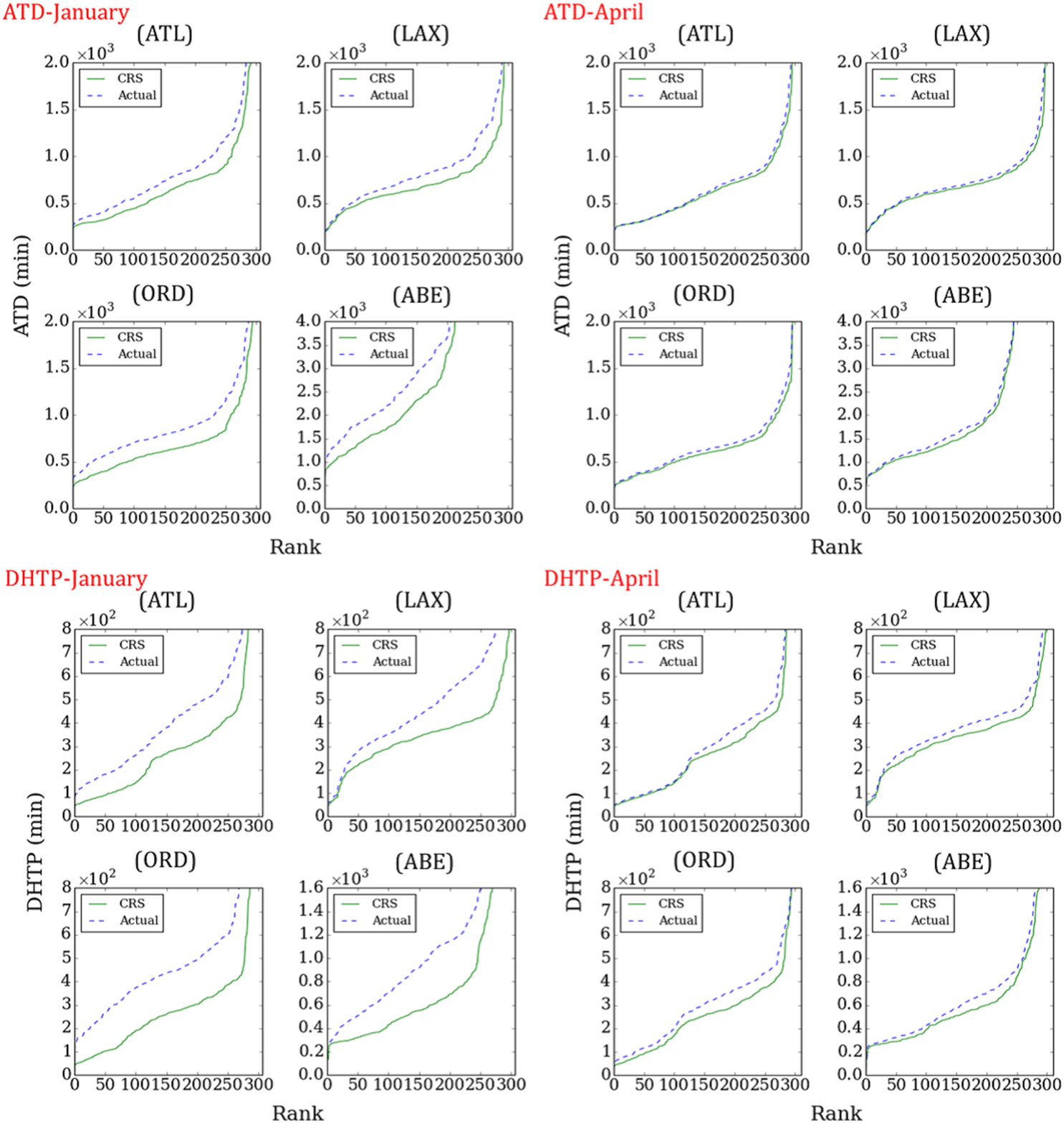}
\caption{(Color online) ATDs and DHTPs from ATL, LAX, ORD, and ABE to destinations in January and April 2014 in ascending order of the TDIs. Green solid and blue dashed lines indicate the scheduled and actual TDIs, respectively.}
\label{fig:9}
\end{center}
\end{figure*}

\section*{APPENDIX B: CALCULATION OF TEMPORAL DISTANCE}
\renewcommand{\theequation}{B.\arabic{equation}}
\setcounter{equation}{0}

In this Appendix, we discuss the way to calculate temporal distances at an arbitrary travel starting time from origins to destinations as proposed by Pan {\it et al.} \cite{pan2011path}. The temporal distance depends on the travel starting time, origin, and destination. Passengers take one or more flights to reach their destinations. The former and latter are a direct flight and flights with connections, respectively. The temporal distance at an arbitrary travel starting time can be calculated by using a list of all shortest paths' departure and arrival times \cite{pan2011path}. Thus, the shortest paths of all combinations of origins and destinations need to be calculated.

The algorithm adopts the notion of vector clocks \cite{mattern1989virtual, fidge1987timestamps}. First, the data with departure and arrival times, origins, and destinations in the dataset are sorted in reverse time order. In addition, we consider the latest arrival times of all combinations of reaching destinations from origins using read data. They are set as infinity for initialization before reading data. Next, the first flight data are read and the latest arrival times are updated. In addition, a search is made for cases in which passengers reach their destinations using this flight as the first flight of their paths. Then, the next data are read and the process is repeated. All the shortest paths are calculated by repeating the process until the last data are read \cite{pan2011path}.

\section*{APPENDIX C: RELATIONSHIP BETWEEN TEMPORAL DISTANCE AND CONNECTION PARAMETERS }
\renewcommand{\theequation}{C.\arabic{equation}}
\setcounter{equation}{0}

\subsection*{Secure time}

Passengers secure at least 45 min for connection in the main sections, which we call ``secure time'' in this paper. However, the length of time recommended for passengers to secure a connection is an issue of importance for construction of better air transportation systems. Thus, we discuss the relationships between the TDIs and secure times in this Appendix. The ATDs and DHTPs are shown as a function of the secure times in Fig~\ref{fig:10}. The left and right graphs are the results with the data recorded in January and April 2014, respectively. Solid, dashed, and dotted lines indicate the scheduled and actual TDIs (left axis) and delays (right axis), respectively. Although the delays in January are larger than in April, the shapes of the functions are similar. The larger the secure times, the larger the scheduled TDIs are. This is because connection times get larger. However, the actual ATDs are almost constant and are minimum when the secure times are $<$34 min (January) or $<$40 min (April). Since delays decrease as the secure times increase, it is better to have these secure times in terms of the ATDs. In addition, the DHTPs have minima at 52-min (January) and 40-min (April) secure times. It is better to have these secure times in terms of the DHTPs. When we argue about how long passengers need to secure a connection, it is necessary to take the decrease of delays in increasing the connection time into consideration. Thus, the appropriate connection time will be slightly larger than the times of minimum ATDs or DHTPs.

\subsection*{Take time}
Passengers take 20 min to transfer to connection planes from arriving ones in the main sections, which we call ``take time'' in this paper. How long it takes to move to connection planes is determined by passengers' walking speed, locations of connection and arriving planes, and airport facilities. This analysis contributes to how important is the impact of optimization of these factors. The ATDs and DHTPs are shown as a function of the take times in Fig~\ref{fig:21}. The left and right graphs show the results in January and April 2014, respectively. Like the previous figure, solid, dashed, and dotted lines indicate the scheduled and actual TDIs (left axis) and delays (right axis), respectively. Scheduled TDIs are constant since scheduled plans are independent of the take times. Actual TDIs and delays monotonically increase as the take times increase because missed connections are likely to occur. The actual TDIs and delays sharply increase when the take times are large. This is because the probability of short delays gets much higher than that of large delays, as shown in Fig.~\ref{fig:4} in Appendix D. Missed connections occur when delays are larger than connection times minus the take times. When the take times are high, even short delays cause missed connections. Thus, large take times makes the probability of missed connections high.

\section*{APPENDIX D: DISTRIBUTION OF DELAYS AND THE NUMBER OF CONNECTIONS}

In this Appendix, we discuss the relationship between the distributions of delays and the number of connections. Figure~\ref{fig:4} shows the (a) PDFs and (b) complementary cumulative distribution functions (CCDFs) of delays. The red solid, blue dotted, and green dashed lines indicate plans with no, one, and two plane connections, respectively. The peak of the delay probability at 1440 min indicates that passengers are put on the next flight the day after their schedules. The probability of delays of $>$100 min significantly increases when passengers have to make plane connections, which is caused by missed connections.

\section*{APPENDIX E: NONPENALTY OF MISSED CONNECTIONS AND SHUFFLED FLIGHT SCHEDULES}
\renewcommand{\theequation}{E.\arabic{equation}}
\setcounter{equation}{0}

Figure~\ref{fig:17} shows monthly averages of (a) ATDs and (b) DHTPs with two variations in January and April, as shown in Fig.~\ref{fig:16}. The results of two variations are compared with the original data (the first and fourth bars). The first variation is nonpenalty (NP). Although we assume that it take 60 min to search for a new flight when passegners miss connections, it does not take any time in the case of nonpenalty. The second variation is shuffled (S), in which case the departure times of flights are shuffled with other flights. Although the TDIs in the case of nonpenalty are smaller than the originals, the differences are not large. This indicates that the time to search for a new flight is not a main factor in causing the delays. In addition, the scheduled TDI in the shuffled case is larger than that in the original, which shows that the flight schedules are designed to facilitate connections. 

Figure~\ref{fig:12} shows the differences in the (a) ATDs and (b) DHTPs between the original and regularized flight schedules with variations in January and April, as shown in Fig.~\ref{fig:11}. The results for the original and nonpenalty cases are similar. In addition to the results for monthly averages of the TDIs, these results also indicate that the time taken searching for a new flight is not an important factor. The TDIs of the shuffled case are increased compared with those of the original data since transit facilitation with burstiness is effective only with well-designed flight schedules. 

Figure~\ref{fig:14} shows the differences in the delays of (a) ATDs and (b) DHTPs between the original and regularized flight schedules with variations in January and April, as shown in Fig.~\ref{fig:13}. The delays for the nonpenalty case in January are increased compared with those of the original data since the delays decrease as transit facilitation is reduced. The delays of the ATD with shuffled data are also increased while those of the DHTP in April are slightly decreased.

\section*{APPENDIX F: FLIGHT SCHEDULES AND ON-TIME PERFORMANCE IN MAJOR AIRPORTS}
\renewcommand{\theequation}{F.\arabic{equation}}
\setcounter{equation}{0}

In this Appendix, we show various analysis results of flight schedules in some airports obtained by using the proposed method. We mainly analyze data from Hartsfield-Jackson Atlanta International Airport (ATL), Los Angeles International Airport (LAX), O'Hare International Airport (ORD), and Lehigh Valley International Airport (ABE).

\subsection*{Dependence of on-time performance on connection times}

In this section, we discuss the dependence of on-time performance on the connection times. The relationships between the connection times and the (a) number of flights, (b) average delay, (c) missed connection rate, and (d) average delays in the missed connections in three main airports (ATL, LAX, and ORD) and all U.S. airports (aggregated data) in January 2014 are shown in Fig.~\ref{fig:6}. The inset of Fig.~\ref{fig:6}(a) shows the difference in the number of flights between the original and randomized data. Although the data are regularized in the main sections for better suitability to the real situation, we implement randomization to the data from the point of view of temporal networks. The randomized data are processed in the same way as the regularization while the randomized scheduled arrival time of the $i$th flights in 2-hour flight schedules is given by 
\begin{equation}
T_i = T_{min} + X_i (T_{max} - T_{min}),
\end{equation}
where $n$ is the number of arrival flights in the 2-hour flight schedules and $X_i$ is the $i$th smallest value of $n$ evenly distributed random numbers greater than or equal to 0 and less than 1. Yellow solid and red dashed lines indicate the original data and a 9-min central moving average, respectively. The number of flights have peaks every 5 min since the scheduled departure and arrival times are often multiples of five. In addition, the number of flights almost exponentially decays since the inter-departure (arrival) time follows an exponential function if we assume that the departure (arrival) times are randomly distributed. However, the numbers of flights with short and long connection times are smaller and larger than the exponential decay, respectively. This indicates that burstiness makes the connection times shorter than the random departures and arrivals. The average delay decreases as the connection time increases while the degree of the trend depends on the airport and the delay increases when the connection time is $\gtrsim$150 min in ORD and all aggregated data. The reasons for the decrease and increase of the delay are a drop in the missed connection rate and a rise in the delays when passengers fail to make connections, respectively. The reason for the latter is that long connection times indicate that the number of operated flights to the destination is small. The passengers are forced to wait a long time for the next flight to the same destination. The balance between the two effects determines whether the average delay decreases or increases.

\subsection*{Dependence of on-time performance on departure and arrival times}

Next, we discuss the dependence of on-time performance on the departure and arrival times. The relationships between the departure time and the (a) number of flights, (b) average delay, (c) missed connection rate, and (d) average delays in the case of the missed connections in three main airports and all U.S. airports (aggregated data) in January 2014 are shown in Fig.~\ref{fig:5}. In addition, the relationships between the arrival time and the (a) number of flights, (b) average delay, (c) missed connection rate, and (d) the average delays in the case of the missed connections in three main airports and all U.S. airports (aggregated data) in January 2014 are shown in Fig.~\ref{fig:7}. The results of the same analysis with the randomized aggregated data from all U.S. airports are shown in Fig.~\ref{fig:8}. Yellow solid and red dashed lines indicate the original data and a 29-min central moving average, respectively. The number of flights of both Fig.~\ref{fig:5} and Fig.~\ref{fig:7} have some peaks, indicating the existence of burstiness. The average delays are larger in the evening. The reason for this is delay propagation \cite{jetzki2009propagation, fleurquin2013systemic}. Delay propagation is a phenomenon in which delays caused by earlier flights using the same aircraft lead to later flights delayed because of the short maintenance times. This leads to severe delays and missed connections. The other reason is the large waiting times for the next flight in the case of missed connections. Since few flights depart and arrive at night, passengers failing transit have to stay at airports overnight, which contributes to very large delays. These characteristics are preserved even if the data are randomized.

\subsection*{Relationship between temporal distance and destinations}

Finally, we discuss the comparison of the TDIs to reach various destinations. In Fig~\ref{fig:9}, the ATDs and DHTPs from three major airports (ATL, LAX, and ORD) and a non-major airport (ABE) to destinations in January and April 2014 are shown in ascending order of the TDIs. Green solid and blue dashed lines indicate the scheduled and actual TDIs, respectively. The numbers of operated airports in January and April are 300 and 304, respectively. Although the scheduled TDIs are similar in January and April, the actual TDIs are longer in January than in April owing to severe delays and a high cancellation rate. The curves for ATL, ORD, and ABE roughly have a downward convex shape whereas these for LAX seem to have two inflection points. The TDIs from major airports to only dozens of destinations are considerably higher, which indicates that hub airports are connected to all airports except them in terms of the TDI. The shape of the curve is similar even in ABE, although the TDI is longer than those of major airports.


\begin{thebibliography}{55}%
\makeatletter
\providecommand \@ifxundefined [1]{%
 \@ifx{#1\undefined}
}%
\providecommand \@ifnum [1]{%
 \ifnum #1\expandafter \@firstoftwo
 \else \expandafter \@secondoftwo
 \fi
}%
\providecommand \@ifx [1]{%
 \ifx #1\expandafter \@firstoftwo
 \else \expandafter \@secondoftwo
 \fi
}%
\providecommand \natexlab [1]{#1}%
\providecommand \enquote  [1]{``#1''}%
\providecommand \bibnamefont  [1]{#1}%
\providecommand \bibfnamefont [1]{#1}%
\providecommand \citenamefont [1]{#1}%
\providecommand \href@noop [0]{\@secondoftwo}%
\providecommand \href [0]{\begingroup \@sanitize@url \@href}%
\providecommand \@href[1]{\@@startlink{#1}\@@href}%
\providecommand \@@href[1]{\endgroup#1\@@endlink}%
\providecommand \@sanitize@url [0]{\catcode `\\12\catcode `\$12\catcode
  `\&12\catcode `\#12\catcode `\^12\catcode `\_12\catcode `\%12\relax}%
\providecommand \@@startlink[1]{}%
\providecommand \@@endlink[0]{}%
\providecommand \url  [0]{\begingroup\@sanitize@url \@url }%
\providecommand \@url [1]{\endgroup\@href {#1}{\urlprefix }}%
\providecommand \urlprefix  [0]{URL }%
\providecommand \Eprint [0]{\href }%
\providecommand \doibase [0]{http://dx.doi.org/}%
\providecommand \selectlanguage [0]{\@gobble}%
\providecommand \bibinfo  [0]{\@secondoftwo}%
\providecommand \bibfield  [0]{\@secondoftwo}%
\providecommand \translation [1]{[#1]}%
\providecommand \BibitemOpen [0]{}%
\providecommand \bibitemStop [0]{}%
\providecommand \bibitemNoStop [0]{.\EOS\space}%
\providecommand \EOS [0]{\spacefactor3000\relax}%
\providecommand \BibitemShut  [1]{\csname bibitem#1\endcsname}%
\let\auto@bib@innerbib\@empty
\bibitem [{\citenamefont {Albert}\ and\ \citenamefont
  {Barab\'asi}(2002)}]{RevModPhys.74.47}%
  \BibitemOpen
  \bibfield  {author} {\bibinfo {author} {\bibfnamefont {R.}~\bibnamefont
  {Albert}}\ and\ \bibinfo {author} {\bibfnamefont {A.-L.}\ \bibnamefont
  {Barab\'asi}},\ }\href {\doibase 10.1103/RevModPhys.74.47} {\bibfield
  {journal} {\bibinfo  {journal} {Rev. Mod. Phys.}\ }\textbf {\bibinfo {volume}
  {74}},\ \bibinfo {pages} {47} (\bibinfo {year} {2002})}\BibitemShut {NoStop}%
\bibitem [{\citenamefont {Castellano}\ \emph {et~al.}(2009)\citenamefont
  {Castellano}, \citenamefont {Fortunato},\ and\ \citenamefont
  {Loreto}}]{RevModPhys.81.591}%
  \BibitemOpen
  \bibfield  {author} {\bibinfo {author} {\bibfnamefont {C.}~\bibnamefont
  {Castellano}}, \bibinfo {author} {\bibfnamefont {S.}~\bibnamefont
  {Fortunato}}, \ and\ \bibinfo {author} {\bibfnamefont {V.}~\bibnamefont
  {Loreto}},\ }\href {\doibase 10.1103/RevModPhys.81.591} {\bibfield  {journal}
  {\bibinfo  {journal} {Rev. Mod. Phys.}\ }\textbf {\bibinfo {volume} {81}},\
  \bibinfo {pages} {591} (\bibinfo {year} {2009})}\BibitemShut {NoStop}%
\bibitem [{\citenamefont {Arenas}\ \emph {et~al.}(2008)\citenamefont {Arenas},
  \citenamefont {D{\'\i}az-Guilera}, \citenamefont {Kurths}, \citenamefont
  {Moreno},\ and\ \citenamefont {Zhou}}]{arenas2008synchronization}%
  \BibitemOpen
  \bibfield  {author} {\bibinfo {author} {\bibfnamefont {A.}~\bibnamefont
  {Arenas}}, \bibinfo {author} {\bibfnamefont {A.}~\bibnamefont
  {D{\'\i}az-Guilera}}, \bibinfo {author} {\bibfnamefont {J.}~\bibnamefont
  {Kurths}}, \bibinfo {author} {\bibfnamefont {Y.}~\bibnamefont {Moreno}}, \
  and\ \bibinfo {author} {\bibfnamefont {C.}~\bibnamefont {Zhou}},\ }\href@noop
  {} {\bibfield  {journal} {\bibinfo  {journal} {Phys. Rep.}\ }\textbf
  {\bibinfo {volume} {469}},\ \bibinfo {pages} {93} (\bibinfo {year}
  {2008})}\BibitemShut {NoStop}%
\bibitem [{\citenamefont {Barth{\'e}lemy}(2011)}]{barthelemy2011spatial}%
  \BibitemOpen
  \bibfield  {author} {\bibinfo {author} {\bibfnamefont {M.}~\bibnamefont
  {Barth{\'e}lemy}},\ }\href@noop {} {\bibfield  {journal} {\bibinfo  {journal}
  {Phys. Rep.}\ }\textbf {\bibinfo {volume} {499}},\ \bibinfo {pages} {1}
  (\bibinfo {year} {2011})}\BibitemShut {NoStop}%
\bibitem [{\citenamefont {Kivel{\"a}}\ \emph {et~al.}(2014)\citenamefont
  {Kivel{\"a}}, \citenamefont {Arenas}, \citenamefont {Barthelemy},
  \citenamefont {Gleeson}, \citenamefont {Moreno},\ and\ \citenamefont
  {Porter}}]{kivela2014multilayer}%
  \BibitemOpen
  \bibfield  {author} {\bibinfo {author} {\bibfnamefont {M.}~\bibnamefont
  {Kivel{\"a}}}, \bibinfo {author} {\bibfnamefont {A.}~\bibnamefont {Arenas}},
  \bibinfo {author} {\bibfnamefont {M.}~\bibnamefont {Barthelemy}}, \bibinfo
  {author} {\bibfnamefont {J.~P.}\ \bibnamefont {Gleeson}}, \bibinfo {author}
  {\bibfnamefont {Y.}~\bibnamefont {Moreno}}, \ and\ \bibinfo {author}
  {\bibfnamefont {M.~A.}\ \bibnamefont {Porter}},\ }\href@noop {} {\bibfield
  {journal} {\bibinfo  {journal} {J. Comp. Netw.}\ }\textbf
  {\bibinfo {volume} {2}},\ \bibinfo {pages} {203} (\bibinfo {year}
  {2014})}\BibitemShut {NoStop}%
\bibitem [{\citenamefont {Pastor-Satorras}\ \emph {et~al.}(2015)\citenamefont
  {Pastor-Satorras}, \citenamefont {Castellano}, \citenamefont {Van~Mieghem},\
  and\ \citenamefont {Vespignani}}]{pastor2015epidemic}%
  \BibitemOpen
  \bibfield  {author} {\bibinfo {author} {\bibfnamefont {R.}~\bibnamefont
  {Pastor-Satorras}}, \bibinfo {author} {\bibfnamefont {C.}~\bibnamefont
  {Castellano}}, \bibinfo {author} {\bibfnamefont {P.}~\bibnamefont
  {Van~Mieghem}}, \ and\ \bibinfo {author} {\bibfnamefont {A.}~\bibnamefont
  {Vespignani}},\ }\href@noop {} {\bibfield  {journal} {\bibinfo  {journal}
  {Rev. Mod. Phys.}\ }\textbf {\bibinfo {volume} {87}},\ \bibinfo
  {pages} {925} (\bibinfo {year} {2015})}\BibitemShut {NoStop}%
\bibitem [{\citenamefont {Newman}(2003)}]{newman2003structure}%
  \BibitemOpen
  \bibfield  {author} {\bibinfo {author} {\bibfnamefont {M.~E.}\ \bibnamefont
  {Newman}},\ }\href@noop {} {\bibfield  {journal} {\bibinfo  {journal} {SIAM
  Rev.}\ }\textbf {\bibinfo {volume} {45}},\ \bibinfo {pages} {167} (\bibinfo
  {year} {2003})}\BibitemShut {NoStop}%
\bibitem [{\citenamefont {Boccaletti}\ \emph {et~al.}(2006)\citenamefont
  {Boccaletti}, \citenamefont {Latora}, \citenamefont {Moreno}, \citenamefont
  {Chavez},\ and\ \citenamefont {Hwang}}]{boccaletti2006complex}%
  \BibitemOpen
  \bibfield  {author} {\bibinfo {author} {\bibfnamefont {S.}~\bibnamefont
  {Boccaletti}}, \bibinfo {author} {\bibfnamefont {V.}~\bibnamefont {Latora}},
  \bibinfo {author} {\bibfnamefont {Y.}~\bibnamefont {Moreno}}, \bibinfo
  {author} {\bibfnamefont {M.}~\bibnamefont {Chavez}}, \ and\ \bibinfo {author}
  {\bibfnamefont {D.-U.}\ \bibnamefont {Hwang}},\ }\href@noop {} {\bibfield
  {journal} {\bibinfo  {journal} {Phys. Rep.}\ }\textbf {\bibinfo {volume}
  {424}},\ \bibinfo {pages} {175} (\bibinfo {year} {2006})}\BibitemShut
  {NoStop}%
\bibitem [{\citenamefont {Holme}\ and\ \citenamefont
  {Saram{\"a}ki}(2012)}]{holme2012temporal}%
  \BibitemOpen
  \bibfield  {author} {\bibinfo {author} {\bibfnamefont {P.}~\bibnamefont
  {Holme}}\ and\ \bibinfo {author} {\bibfnamefont {J.}~\bibnamefont
  {Saram{\"a}ki}},\ }\href@noop {} {\bibfield  {journal} {\bibinfo  {journal}
  {Phys. Rep.}\ }\textbf {\bibinfo {volume} {519}},\ \bibinfo {pages} {97}
  (\bibinfo {year} {2012})}\BibitemShut {NoStop}%
\bibitem [{\citenamefont {Eckmann}\ \emph {et~al.}(2004)\citenamefont
  {Eckmann}, \citenamefont {Moses},\ and\ \citenamefont
  {Sergi}}]{eckmann2004entropy}%
  \BibitemOpen
  \bibfield  {author} {\bibinfo {author} {\bibfnamefont {J.-P.}\ \bibnamefont
  {Eckmann}}, \bibinfo {author} {\bibfnamefont {E.}~\bibnamefont {Moses}}, \
  and\ \bibinfo {author} {\bibfnamefont {D.}~\bibnamefont {Sergi}},\
  }\href@noop {} {\bibfield  {journal} {\bibinfo  {journal} {Proc. Natl. Acad.
  Sci. USA}\ }\textbf {\bibinfo {volume} {101}},\ \bibinfo {pages} {14333}
  (\bibinfo {year} {2004})}\BibitemShut {NoStop}%
\bibitem [{\citenamefont {Barab{\'a}si}(2005)}]{barabasi2005origin}%
  \BibitemOpen
  \bibfield  {author} {\bibinfo {author} {\bibfnamefont {A.-L.}\ \bibnamefont
  {Barab{\'a}si}},\ }\href@noop {} {\bibfield  {journal} {\bibinfo  {journal}
  {Nature (London)}\ }\textbf {\bibinfo {volume} {435}},\ \bibinfo {pages}
  {207} (\bibinfo {year} {2005})}\BibitemShut {NoStop}%
\bibitem [{\citenamefont {Karsai}\ \emph {et~al.}(2012)\citenamefont {Karsai},
  \citenamefont {Kaski}, \citenamefont {Barab{\'a}si},\ and\ \citenamefont
  {Kert{\'e}sz}}]{karsai2012universal}%
  \BibitemOpen
  \bibfield  {author} {\bibinfo {author} {\bibfnamefont {M.}~\bibnamefont
  {Karsai}}, \bibinfo {author} {\bibfnamefont {K.}~\bibnamefont {Kaski}},
  \bibinfo {author} {\bibfnamefont {A.-L.}\ \bibnamefont {Barab{\'a}si}}, \
  and\ \bibinfo {author} {\bibfnamefont {J.}~\bibnamefont {Kert{\'e}sz}},\
  }\href@noop {} {\bibfield  {journal} {\bibinfo  {journal} {Sci. Rep.}\
  }\textbf {\bibinfo {volume} {2}} (\bibinfo {year} {2012})}\BibitemShut
  {NoStop}%
\bibitem [{\citenamefont {V{\'a}zquez}\ \emph {et~al.}(2006)\citenamefont
  {V{\'a}zquez}, \citenamefont {Oliveira}, \citenamefont {Dezs{\"o}},
  \citenamefont {Goh}, \citenamefont {Kondor},\ and\ \citenamefont
  {Barab{\'a}si}}]{vazquez2006modeling}%
  \BibitemOpen
  \bibfield  {author} {\bibinfo {author} {\bibfnamefont {A.}~\bibnamefont
  {V{\'a}zquez}}, \bibinfo {author} {\bibfnamefont {J.~G.}\ \bibnamefont
  {Oliveira}}, \bibinfo {author} {\bibfnamefont {Z.}~\bibnamefont {Dezs{\"o}}},
  \bibinfo {author} {\bibfnamefont {K.-I.}\ \bibnamefont {Goh}}, \bibinfo
  {author} {\bibfnamefont {I.}~\bibnamefont {Kondor}}, \ and\ \bibinfo {author}
  {\bibfnamefont {A.-L.}\ \bibnamefont {Barab{\'a}si}},\ }\href@noop {}
  {\bibfield  {journal} {\bibinfo  {journal} {Phys. Rev. E}\ }\textbf {\bibinfo
  {volume} {73}},\ \bibinfo {pages} {036127} (\bibinfo {year}
  {2006})}\BibitemShut {NoStop}%
\bibitem [{\citenamefont {Goh}\ and\ \citenamefont
  {Barab{\'a}si}(2008)}]{goh2008burstiness}%
  \BibitemOpen
  \bibfield  {author} {\bibinfo {author} {\bibfnamefont {K.-I.}\ \bibnamefont
  {Goh}}\ and\ \bibinfo {author} {\bibfnamefont {A.-L.}\ \bibnamefont
  {Barab{\'a}si}},\ }\href@noop {} {\bibfield  {journal} {\bibinfo  {journal}
  {EPL}\ }\textbf {\bibinfo {volume} {81}},\ \bibinfo {pages} {48002} (\bibinfo
  {year} {2008})}\BibitemShut {NoStop}%
\bibitem [{\citenamefont {Cattuto}\ \emph {et~al.}(2010)\citenamefont
  {Cattuto}, \citenamefont {Van~den Broeck}, \citenamefont {Barrat},
  \citenamefont {Colizza}, \citenamefont {Pinton},\ and\ \citenamefont
  {Vespignani}}]{cattuto2010dynamics}%
  \BibitemOpen
  \bibfield  {author} {\bibinfo {author} {\bibfnamefont {C.}~\bibnamefont
  {Cattuto}}, \bibinfo {author} {\bibfnamefont {W.}~\bibnamefont {Van~den
  Broeck}}, \bibinfo {author} {\bibfnamefont {A.}~\bibnamefont {Barrat}},
  \bibinfo {author} {\bibfnamefont {V.}~\bibnamefont {Colizza}}, \bibinfo
  {author} {\bibfnamefont {J.-F.}\ \bibnamefont {Pinton}}, \ and\ \bibinfo
  {author} {\bibfnamefont {A.}~\bibnamefont {Vespignani}},\ }\href@noop {}
  {\bibfield  {journal} {\bibinfo  {journal} {PloS ONE}\ }\textbf {\bibinfo
  {volume} {5}},\ \bibinfo {pages} {e11596} (\bibinfo {year}
  {2010})}\BibitemShut {NoStop}%
\bibitem [{\citenamefont {Kemuriyama}\ \emph {et~al.}(2010)\citenamefont
  {Kemuriyama}, \citenamefont {Ohta}, \citenamefont {Sato}, \citenamefont
  {Maruyama}, \citenamefont {Tandai-Hiruma}, \citenamefont {Kato},\ and\
  \citenamefont {Nishida}}]{kemuriyama2010power}%
  \BibitemOpen
  \bibfield  {author} {\bibinfo {author} {\bibfnamefont {T.}~\bibnamefont
  {Kemuriyama}}, \bibinfo {author} {\bibfnamefont {H.}~\bibnamefont {Ohta}},
  \bibinfo {author} {\bibfnamefont {Y.}~\bibnamefont {Sato}}, \bibinfo {author}
  {\bibfnamefont {S.}~\bibnamefont {Maruyama}}, \bibinfo {author}
  {\bibfnamefont {M.}~\bibnamefont {Tandai-Hiruma}}, \bibinfo {author}
  {\bibfnamefont {K.}~\bibnamefont {Kato}}, \ and\ \bibinfo {author}
  {\bibfnamefont {Y.}~\bibnamefont {Nishida}},\ }\href@noop {} {\bibfield
  {journal} {\bibinfo  {journal} {Biolog. Syst.}\ }\textbf {\bibinfo {volume}
  {101}},\ \bibinfo {pages} {144} (\bibinfo {year} {2010})}\BibitemShut
  {NoStop}%
\bibitem [{\citenamefont {Saichev}\ and\ \citenamefont
  {Sornette}(2006)}]{saichev2006universal}%
  \BibitemOpen
  \bibfield  {author} {\bibinfo {author} {\bibfnamefont {A.}~\bibnamefont
  {Saichev}}\ and\ \bibinfo {author} {\bibfnamefont {D.}~\bibnamefont
  {Sornette}},\ }\href@noop {} {\bibfield  {journal} {\bibinfo  {journal}
  {Phys. Rev. Lett.}\ }\textbf {\bibinfo {volume} {97}},\ \bibinfo {pages}
  {078501} (\bibinfo {year} {2006})}\BibitemShut {NoStop}%
\bibitem [{\citenamefont {Karsai}\ \emph {et~al.}(2011)\citenamefont {Karsai},
  \citenamefont {Kivel{\"a}}, \citenamefont {Pan}, \citenamefont {Kaski},
  \citenamefont {Kert{\'e}sz}, \citenamefont {Barab{\'a}si},\ and\
  \citenamefont {Saram{\"a}ki}}]{karsai2011small}%
  \BibitemOpen
  \bibfield  {author} {\bibinfo {author} {\bibfnamefont {M.}~\bibnamefont
  {Karsai}}, \bibinfo {author} {\bibfnamefont {M.}~\bibnamefont {Kivel{\"a}}},
  \bibinfo {author} {\bibfnamefont {R.~K.}\ \bibnamefont {Pan}}, \bibinfo
  {author} {\bibfnamefont {K.}~\bibnamefont {Kaski}}, \bibinfo {author}
  {\bibfnamefont {J.}~\bibnamefont {Kert{\'e}sz}}, \bibinfo {author}
  {\bibfnamefont {A.-L.}\ \bibnamefont {Barab{\'a}si}}, \ and\ \bibinfo
  {author} {\bibfnamefont {J.}~\bibnamefont {Saram{\"a}ki}},\ }\href@noop {}
  {\bibfield  {journal} {\bibinfo  {journal} {Phys. Rev. E}\ }\textbf {\bibinfo
  {volume} {83}},\ \bibinfo {pages} {025102} (\bibinfo {year}
  {2011})}\BibitemShut {NoStop}%
\bibitem [{\citenamefont {Iribarren}\ and\ \citenamefont
  {Moro}(2009)}]{iribarren2009impact}%
  \BibitemOpen
  \bibfield  {author} {\bibinfo {author} {\bibfnamefont {J.~L.}\ \bibnamefont
  {Iribarren}}\ and\ \bibinfo {author} {\bibfnamefont {E.}~\bibnamefont
  {Moro}},\ }\href@noop {} {\bibfield  {journal} {\bibinfo  {journal} {Phys.
  Rev. Lett.}\ }\textbf {\bibinfo {volume} {103}},\ \bibinfo {pages} {038702}
  (\bibinfo {year} {2009})}\BibitemShut {NoStop}%
\bibitem [{\citenamefont {Vazquez}\ \emph {et~al.}(2007)\citenamefont
  {Vazquez}, \citenamefont {Racz}, \citenamefont {Lukacs},\ and\ \citenamefont
  {Barab{\'a}si}}]{vazquez2007impact2}%
  \BibitemOpen
  \bibfield  {author} {\bibinfo {author} {\bibfnamefont {A.}~\bibnamefont
  {Vazquez}}, \bibinfo {author} {\bibfnamefont {B.}~\bibnamefont {Racz}},
  \bibinfo {author} {\bibfnamefont {A.}~\bibnamefont {Lukacs}}, \ and\ \bibinfo
  {author} {\bibfnamefont {A.-L.}\ \bibnamefont {Barab{\'a}si}},\ }\href@noop
  {} {\bibfield  {journal} {\bibinfo  {journal} {Phys. Rev. Lett.}\ }\textbf
  {\bibinfo {volume} {98}},\ \bibinfo {pages} {158702} (\bibinfo {year}
  {2007})}\BibitemShut {NoStop}%
\bibitem [{\citenamefont {Jo}\ \emph {et~al.}(2014)\citenamefont {Jo},
  \citenamefont {Perotti}, \citenamefont {Kaski},\ and\ \citenamefont
  {Kert{\'e}sz}}]{jo2014analytically}%
  \BibitemOpen
  \bibfield  {author} {\bibinfo {author} {\bibfnamefont {H.-H.}\ \bibnamefont
  {Jo}}, \bibinfo {author} {\bibfnamefont {J.~I.}\ \bibnamefont {Perotti}},
  \bibinfo {author} {\bibfnamefont {K.}~\bibnamefont {Kaski}}, \ and\ \bibinfo
  {author} {\bibfnamefont {J.}~\bibnamefont {Kert{\'e}sz}},\ }\href@noop {}
  {\bibfield  {journal} {\bibinfo  {journal} {Phys. Rev. X}\ }\textbf {\bibinfo
  {volume} {4}},\ \bibinfo {pages} {011041} (\bibinfo {year}
  {2014})}\BibitemShut {NoStop}%
\bibitem [{\citenamefont {Gavald{\`a}-Miralles}\ \emph
  {et~al.}(2014)\citenamefont {Gavald{\`a}-Miralles}, \citenamefont {Choffnes},
  \citenamefont {Otto}, \citenamefont {S{\'a}nchez}, \citenamefont
  {Bustamante}, \citenamefont {Amaral}, \citenamefont {Duch},\ and\
  \citenamefont {Guimer{\`a}}}]{gavalda2014impact}%
  \BibitemOpen
  \bibfield  {author} {\bibinfo {author} {\bibfnamefont {A.}~\bibnamefont
  {Gavald{\`a}-Miralles}}, \bibinfo {author} {\bibfnamefont {D.~R.}\
  \bibnamefont {Choffnes}}, \bibinfo {author} {\bibfnamefont {J.~S.}\
  \bibnamefont {Otto}}, \bibinfo {author} {\bibfnamefont {M.~A.}\ \bibnamefont
  {S{\'a}nchez}}, \bibinfo {author} {\bibfnamefont {F.~E.}\ \bibnamefont
  {Bustamante}}, \bibinfo {author} {\bibfnamefont {L.~A.}\ \bibnamefont
  {Amaral}}, \bibinfo {author} {\bibfnamefont {J.}~\bibnamefont {Duch}}, \ and\
  \bibinfo {author} {\bibfnamefont {R.}~\bibnamefont {Guimer{\`a}}},\
  }\href@noop {} {\bibfield  {journal} {\bibinfo  {journal} {Proc. Natl. Acad.
  Sci. USA}\ }\textbf {\bibinfo {volume} {111}},\ \bibinfo {pages} {15322}
  (\bibinfo {year} {2014})}\BibitemShut {NoStop}%
\bibitem [{\citenamefont {Horv{\'a}th}\ and\ \citenamefont
  {Kert{\'e}sz}(2014)}]{horvath2014spreading}%
  \BibitemOpen
  \bibfield  {author} {\bibinfo {author} {\bibfnamefont {D.~X.}\ \bibnamefont
  {Horv{\'a}th}}\ and\ \bibinfo {author} {\bibfnamefont {J.}~\bibnamefont
  {Kert{\'e}sz}},\ }\href@noop {} {\bibfield  {journal} {\bibinfo  {journal}
  {New J. Phys.}\ }\textbf {\bibinfo {volume} {16}},\ \bibinfo {pages} {073037}
  (\bibinfo {year} {2014})}\BibitemShut {NoStop}%
\bibitem [{\citenamefont {Moinet}\ \emph {et~al.}(2015)\citenamefont {Moinet},
  \citenamefont {Starnini},\ and\ \citenamefont
  {Pastor-Satorras}}]{PhysRevLett.114.108701}%
  \BibitemOpen
  \bibfield  {author} {\bibinfo {author} {\bibfnamefont {A.}~\bibnamefont
  {Moinet}}, \bibinfo {author} {\bibfnamefont {M.}~\bibnamefont {Starnini}}, \
  and\ \bibinfo {author} {\bibfnamefont {R.}~\bibnamefont {Pastor-Satorras}},\
  }\href {\doibase 10.1103/PhysRevLett.114.108701} {\bibfield  {journal}
  {\bibinfo  {journal} {Phys. Rev. Lett.}\ }\textbf {\bibinfo {volume} {114}},\
  \bibinfo {pages} {108701} (\bibinfo {year} {2015})}\BibitemShut {NoStop}%
\bibitem [{\citenamefont {Ito}\ and\ \citenamefont
  {Nishinari}(2015)}]{Ito2015universal}%
  \BibitemOpen
  \bibfield  {author} {\bibinfo {author} {\bibfnamefont {H.}~\bibnamefont
  {Ito}}\ and\ \bibinfo {author} {\bibfnamefont {K.}~\bibnamefont
  {Nishinari}},\ }\href {\doibase 10.1103/PhysRevE.92.062815} {\bibfield
  {journal} {\bibinfo  {journal} {Phys. Rev. E}\ }\textbf {\bibinfo {volume}
  {92}},\ \bibinfo {pages} {062815} (\bibinfo {year} {2015})}\BibitemShut
  {NoStop}%
\bibitem [{\citenamefont {Peterson}\ \emph {et~al.}(1995)\citenamefont
  {Peterson}, \citenamefont {Bertsimas},\ and\ \citenamefont
  {Odoni}}]{peterson1995models}%
  \BibitemOpen
  \bibfield  {author} {\bibinfo {author} {\bibfnamefont {M.~D.}\ \bibnamefont
  {Peterson}}, \bibinfo {author} {\bibfnamefont {D.~J.}\ \bibnamefont
  {Bertsimas}}, \ and\ \bibinfo {author} {\bibfnamefont {A.~R.}\ \bibnamefont
  {Odoni}},\ }\href@noop {} {\bibfield  {journal} {\bibinfo  {journal}
  {Manag. Sci.}\ }\textbf {\bibinfo {volume} {41}},\ \bibinfo {pages}
  {1279} (\bibinfo {year} {1995})}\BibitemShut {NoStop}%
\bibitem [{\citenamefont {Ater}(2012)}]{ater2012internalization}%
  \BibitemOpen
  \bibfield  {author} {\bibinfo {author} {\bibfnamefont {I.}~\bibnamefont
  {Ater}},\ }\href@noop {} {\bibfield  {journal} {\bibinfo  {journal} {J. Urban. Econ.}\ }\textbf {\bibinfo {volume} {72}},\ \bibinfo {pages}
  {196} (\bibinfo {year} {2012})}\BibitemShut {NoStop}%
\bibitem [{\citenamefont {Gilbo}(1993)}]{gilbo1993airport}%
  \BibitemOpen
  \bibfield  {author} {\bibinfo {author} {\bibfnamefont {E.~P.}\ \bibnamefont
  {Gilbo}},\ }\href@noop {} {\bibfield  {journal} {\bibinfo  {journal} {IEEE Trans. Control Syst. Technol.}\ }\textbf {\bibinfo {volume} {1}},\
  \bibinfo {pages} {144} (\bibinfo {year} {1993})}\BibitemShut {NoStop}%
\bibitem [{\citenamefont {Bootsma}(1997)}]{bootsma1997airline}%
  \BibitemOpen
  \bibfield  {author} {\bibinfo {author} {\bibfnamefont {P.~D.}\ \bibnamefont
  {Bootsma}},\ }\href@noop {} {\bibfield  {journal} {\bibinfo  {journal} {PhD
  diss., University of Twente}\ } (\bibinfo {year} {1997})}\BibitemShut
  {NoStop}%
\bibitem [{\citenamefont {Alderighi}\ \emph {et~al.}(2007)\citenamefont
  {Alderighi}, \citenamefont {Cento}, \citenamefont {Nijkamp},\ and\
  \citenamefont {Rietveld}}]{alderighi2007assessment}%
  \BibitemOpen
  \bibfield  {author} {\bibinfo {author} {\bibfnamefont {M.}~\bibnamefont
  {Alderighi}}, \bibinfo {author} {\bibfnamefont {A.}~\bibnamefont {Cento}},
  \bibinfo {author} {\bibfnamefont {P.}~\bibnamefont {Nijkamp}}, \ and\
  \bibinfo {author} {\bibfnamefont {P.}~\bibnamefont {Rietveld}},\ }\href@noop
  {} {\bibfield  {journal} {\bibinfo  {journal} {Trans. Rev.}\ }\textbf
  {\bibinfo {volume} {27}},\ \bibinfo {pages} {529} (\bibinfo {year}
  {2007})}\BibitemShut {NoStop}%
\bibitem [{\citenamefont {Burghouwt}\ and\ \citenamefont
  {de~Wit}(2005)}]{burghouwt2005temporal}%
  \BibitemOpen
  \bibfield  {author} {\bibinfo {author} {\bibfnamefont {G.}~\bibnamefont
  {Burghouwt}}\ and\ \bibinfo {author} {\bibfnamefont {J.}~\bibnamefont
  {de~Wit}},\ }\href@noop {} {\bibfield  {journal} {\bibinfo  {journal}
  {J. Air Trans. Manag.}\ }\textbf {\bibinfo {volume} {11}},\
  \bibinfo {pages} {185} (\bibinfo {year} {2005})}\BibitemShut {NoStop}%
\bibitem [{\citenamefont {Dennis}(1999)}]{dennis1999competition}%
  \BibitemOpen
  \bibfield  {author} {\bibinfo {author} {\bibfnamefont {N.}~\bibnamefont
  {Dennis}},\ }in\ \href@noop {} {\emph {\bibinfo {booktitle} {proceedings of  8th World Conference on Transport Research}}},\ \bibinfo {series and number} {\bibinfo {number} {Volume 1}}\
  (\bibinfo {year} {World Conference on Transport Research Society, 1999})\BibitemShut {NoStop}%
\bibitem [{\citenamefont {Costa}\ \emph {et~al.}(2011)\citenamefont {Costa},
  \citenamefont {Oliveira~Jr}, \citenamefont {Travieso}, \citenamefont
  {Rodrigues}, \citenamefont {Villas~Boas}, \citenamefont {Antiqueira},
  \citenamefont {Viana},\ and\ \citenamefont
  {Correa~Rocha}}]{costa2011analyzing}%
  \BibitemOpen
  \bibfield  {author} {\bibinfo {author} {\bibfnamefont {L.~d.~F.}\
  \bibnamefont {Costa}}, \bibinfo {author} {\bibfnamefont {O.~N.}\ \bibnamefont
  {Oliveira~Jr}}, \bibinfo {author} {\bibfnamefont {G.}~\bibnamefont
  {Travieso}}, \bibinfo {author} {\bibfnamefont {F.~A.}\ \bibnamefont
  {Rodrigues}}, \bibinfo {author} {\bibfnamefont {P.~R.}\ \bibnamefont
  {Villas~Boas}}, \bibinfo {author} {\bibfnamefont {L.}~\bibnamefont
  {Antiqueira}}, \bibinfo {author} {\bibfnamefont {M.~P.}\ \bibnamefont
  {Viana}}, \ and\ \bibinfo {author} {\bibfnamefont {L.~E.}\ \bibnamefont
  {Correa~Rocha}},\ }\href@noop {} {\bibfield  {journal} {\bibinfo  {journal}
  {?Adv. Phys.}\ }\textbf {\bibinfo {volume} {60}},\ \bibinfo {pages}
  {329} (\bibinfo {year} {2011})}\BibitemShut {NoStop}%
\bibitem [{\citenamefont {Chowdhury}\ \emph {et~al.}(2000)\citenamefont
  {Chowdhury}, \citenamefont {Santen},\ and\ \citenamefont
  {Schadschneider}}]{chowdhury2000statistical}%
  \BibitemOpen
  \bibfield  {author} {\bibinfo {author} {\bibfnamefont {D.}~\bibnamefont
  {Chowdhury}}, \bibinfo {author} {\bibfnamefont {L.}~\bibnamefont {Santen}}, \
  and\ \bibinfo {author} {\bibfnamefont {A.}~\bibnamefont {Schadschneider}},\
  }\href@noop {} {\bibfield  {journal} {\bibinfo  {journal} {Phys. Rep.}\
  }\textbf {\bibinfo {volume} {329}},\ \bibinfo {pages} {199} (\bibinfo {year}
  {2000})}\BibitemShut {NoStop}%
\bibitem [{\citenamefont {Helbing}(2001)}]{helbing2001traffic}%
  \BibitemOpen
  \bibfield  {author} {\bibinfo {author} {\bibfnamefont {D.}~\bibnamefont
  {Helbing}},\ }\href@noop {} {\bibfield  {journal} {\bibinfo  {journal} {Rev.
  Mod. Phys.}\ }\textbf {\bibinfo {volume} {73}},\ \bibinfo {pages} {1067}
  (\bibinfo {year} {2001})}\BibitemShut {NoStop}%
\bibitem [{\citenamefont {Bryan}\ and\ \citenamefont
  {O'Kelly}(1999)}]{bryan1999hub}%
  \BibitemOpen
  \bibfield  {author} {\bibinfo {author} {\bibfnamefont {D.~L.}\ \bibnamefont
  {Bryan}}\ and\ \bibinfo {author} {\bibfnamefont {M.~E.}\ \bibnamefont
  {O'Kelly}},\ }\href@noop {} {\bibfield  {journal} {\bibinfo  {journal} {J.
  Reg. Sci.}\ }\textbf {\bibinfo {volume} {39}},\ \bibinfo {pages} {275}
  (\bibinfo {year} {1999})}\BibitemShut {NoStop}%
\bibitem [{\citenamefont {Guimera}\ \emph {et~al.}(2005)\citenamefont
  {Guimera}, \citenamefont {Mossa}, \citenamefont {Turtschi},\ and\
  \citenamefont {Amaral}}]{guimera2005worldwide}%
  \BibitemOpen
  \bibfield  {author} {\bibinfo {author} {\bibfnamefont {R.}~\bibnamefont
  {Guimera}}, \bibinfo {author} {\bibfnamefont {S.}~\bibnamefont {Mossa}},
  \bibinfo {author} {\bibfnamefont {A.}~\bibnamefont {Turtschi}}, \ and\
  \bibinfo {author} {\bibfnamefont {L.~N.}\ \bibnamefont {Amaral}},\
  }\href@noop {} {\bibfield  {journal} {\bibinfo  {journal} {Proc. Natl. Acad.
  Sci. USA}\ }\textbf {\bibinfo {volume} {102}},\ \bibinfo {pages} {7794}
  (\bibinfo {year} {2005})}\BibitemShut {NoStop}%
\bibitem [{\citenamefont {Karamouzas}\ \emph {et~al.}(2014)\citenamefont
  {Karamouzas}, \citenamefont {Skinner},\ and\ \citenamefont
  {Guy}}]{karamouzas2014universal}%
  \BibitemOpen
  \bibfield  {author} {\bibinfo {author} {\bibfnamefont {I.}~\bibnamefont
  {Karamouzas}}, \bibinfo {author} {\bibfnamefont {B.}~\bibnamefont {Skinner}},
  \ and\ \bibinfo {author} {\bibfnamefont {S.~J.}\ \bibnamefont {Guy}},\
  }\href@noop {} {\bibfield  {journal} {\bibinfo  {journal} {Phys. Rev.
  Lett.}\ }\textbf {\bibinfo {volume} {113}},\ \bibinfo {pages} {238701}
  (\bibinfo {year} {2014})}\BibitemShut {NoStop}%
\bibitem [{\citenamefont {Li}\ \emph {et~al.}(2015)\citenamefont {Li},
  \citenamefont {Fu}, \citenamefont {Wang}, \citenamefont {Lu}, \citenamefont
  {Berezin}, \citenamefont {Stanley},\ and\ \citenamefont
  {Havlin}}]{li2015percolation}%
  \BibitemOpen
  \bibfield  {author} {\bibinfo {author} {\bibfnamefont {D.}~\bibnamefont
  {Li}}, \bibinfo {author} {\bibfnamefont {B.}~\bibnamefont {Fu}}, \bibinfo
  {author} {\bibfnamefont {Y.}~\bibnamefont {Wang}}, \bibinfo {author}
  {\bibfnamefont {G.}~\bibnamefont {Lu}}, \bibinfo {author} {\bibfnamefont
  {Y.}~\bibnamefont {Berezin}}, \bibinfo {author} {\bibfnamefont {H.~E.}\
  \bibnamefont {Stanley}}, \ and\ \bibinfo {author} {\bibfnamefont
  {S.}~\bibnamefont {Havlin}},\ }\href@noop {} {\bibfield  {journal} {\bibinfo
  {journal} {Proc. Natl. Acad. Sci. USA}\ }\textbf {\bibinfo {volume} {112}},\
  \bibinfo {pages} {669} (\bibinfo {year} {2015})}\BibitemShut {NoStop}%
\bibitem [{\citenamefont {Airbus}(2014)}]{Airbus}%
  \BibitemOpen
  \bibfield  {author} {\bibinfo {author} {\bibnamefont {Airbus,}}} {\enquote {\bibinfo {title}
  {Global market forecast 2014-2033}} (http://www.airbus.com) }\BibitemShut
  {NoStop}%
\bibitem [{\citenamefont {Gwiggner}\ and\ \citenamefont
  {Nagaoka}(2014)}]{gwiggner2014data}%
  \BibitemOpen
  \bibfield  {author} {\bibinfo {author} {\bibfnamefont {C.}~\bibnamefont
  {Gwiggner}}\ and\ \bibinfo {author} {\bibfnamefont {S.}~\bibnamefont
  {Nagaoka}},\ }\href@noop {} {\bibfield  {journal} {\bibinfo  {journal} {Eur.
  J. Oper. Res.}\ }\textbf {\bibinfo {volume} {235}},\ \bibinfo {pages} {265}
  (\bibinfo {year} {2014})}\BibitemShut {NoStop}%
\bibitem [{\citenamefont {Lacasa}\ \emph {et~al.}(2009)\citenamefont {Lacasa},
  \citenamefont {Cea},\ and\ \citenamefont {Zanin}}]{lacasa2009jamming}%
  \BibitemOpen
  \bibfield  {author} {\bibinfo {author} {\bibfnamefont {L.}~\bibnamefont
  {Lacasa}}, \bibinfo {author} {\bibfnamefont {M.}~\bibnamefont {Cea}}, \ and\
  \bibinfo {author} {\bibfnamefont {M.}~\bibnamefont {Zanin}},\ }\href@noop {}
  {\bibfield  {journal} {\bibinfo  {journal} {Physica A}\ }\textbf {\bibinfo
  {volume} {388}},\ \bibinfo {pages} {3948} (\bibinfo {year}
  {2009})}\BibitemShut {NoStop}%
\bibitem [{\citenamefont {Ezaki}\ \emph {et~al.}(2015)\citenamefont {Ezaki},
  \citenamefont {Nishi},\ and\ \citenamefont {Nishinari}}]{ezaki2015taming}%
  \BibitemOpen
  \bibfield  {author} {\bibinfo {author} {\bibfnamefont {T.}~\bibnamefont
  {Ezaki}}, \bibinfo {author} {\bibfnamefont {R.}~\bibnamefont {Nishi}}, \ and\
  \bibinfo {author} {\bibfnamefont {K.}~\bibnamefont {Nishinari}},\ }\href@noop
  {} {\bibfield  {journal} {\bibinfo  {journal} {J. Stat. Mech. Theor. Exp.}\ }\textbf {\bibinfo {volume} {2015}},\
  \bibinfo {pages} {P06013} (\bibinfo {year} {2015})}\BibitemShut {NoStop}%
\bibitem [{\citenamefont {of~Transportation Statistics~(BTS)}(2014)}]{BTS}%
  \BibitemOpen
  \bibfield  {author} {\bibinfo {author} {\bibfnamefont {Bureau}~\bibnamefont
  {of~Transportation Statistics~(BTS)}},} {\enquote {\bibinfo {title} {Rita database}} (http://www.transtats.bts.gov) }\BibitemShut {NoStop}%
\bibitem [{\citenamefont {AhmadBeygi}\ \emph {et~al.}(2008)\citenamefont
  {AhmadBeygi}, \citenamefont {Cohn}, \citenamefont {Guan},\ and\ \citenamefont
  {Belobaba}}]{ahmadbeygi2008analysis}%
  \BibitemOpen
  \bibfield  {author} {\bibinfo {author} {\bibfnamefont {S.}~\bibnamefont
  {AhmadBeygi}}, \bibinfo {author} {\bibfnamefont {A.}~\bibnamefont {Cohn}},
  \bibinfo {author} {\bibfnamefont {Y.}~\bibnamefont {Guan}}, \ and\ \bibinfo
  {author} {\bibfnamefont {P.}~\bibnamefont {Belobaba}},\ }\href@noop {}
  {\bibfield  {journal} {\bibinfo  {journal} {J. Air Trans.
  Manag.}\ }\textbf {\bibinfo {volume} {14}},\ \bibinfo {pages} {221}
  (\bibinfo {year} {2008})}\BibitemShut {NoStop}%
\bibitem [{\citenamefont {Ahmadbeygi}\ \emph {et~al.}(2010)\citenamefont
  {Ahmadbeygi}, \citenamefont {Cohn},\ and\ \citenamefont
  {Lapp}}]{ahmadbeygi2010decreasing}%
  \BibitemOpen
  \bibfield  {author} {\bibinfo {author} {\bibfnamefont {S.}~\bibnamefont
  {Ahmadbeygi}}, \bibinfo {author} {\bibfnamefont {A.}~\bibnamefont {Cohn}}, \
  and\ \bibinfo {author} {\bibfnamefont {M.}~\bibnamefont {Lapp}},\ }\href@noop
  {} {\bibfield  {journal} {\bibinfo  {journal} {IIE Trans.}\ }\textbf
  {\bibinfo {volume} {42}},\ \bibinfo {pages} {478} (\bibinfo {year}
  {2010})}\BibitemShut {NoStop}%
\bibitem [{\citenamefont {Barnhart}\ \emph {et~al.}(2012)\citenamefont
  {Barnhart}, \citenamefont {Fearing}, \citenamefont {Odoni},\ and\
  \citenamefont {Vaze}}]{barnhart2012demand}%
  \BibitemOpen
  \bibfield  {author} {\bibinfo {author} {\bibfnamefont {C.}~\bibnamefont
  {Barnhart}}, \bibinfo {author} {\bibfnamefont {D.}~\bibnamefont {Fearing}},
  \bibinfo {author} {\bibfnamefont {A.}~\bibnamefont {Odoni}}, \ and\ \bibinfo
  {author} {\bibfnamefont {V.}~\bibnamefont {Vaze}},\ }\href@noop {} {\bibfield
   {journal} {\bibinfo  {journal} {EURO J. Transport. Log.}\ }\textbf {\bibinfo
  {volume} {1}},\ \bibinfo {pages} {135} (\bibinfo {year} {2012})}\BibitemShut
  {NoStop}%
\bibitem [{\citenamefont {Barnhart}\ \emph {et~al.}(2014)\citenamefont
  {Barnhart}, \citenamefont {Fearing},\ and\ \citenamefont
  {Vaze}}]{barnhart2014modeling}%
  \BibitemOpen
  \bibfield  {author} {\bibinfo {author} {\bibfnamefont {C.}~\bibnamefont
  {Barnhart}}, \bibinfo {author} {\bibfnamefont {D.}~\bibnamefont {Fearing}}, \
  and\ \bibinfo {author} {\bibfnamefont {V.}~\bibnamefont {Vaze}},\ }\href@noop
  {} {\bibfield  {journal} {\bibinfo  {journal} {Oper. Res.}\ }\textbf
  {\bibinfo {volume} {62}},\ \bibinfo {pages} {580} (\bibinfo {year}
  {2014})}\BibitemShut {NoStop}%
\bibitem [{\citenamefont {Holme}(2015)}]{holme2015modern}%
  \BibitemOpen
  \bibfield  {author} {\bibinfo {author} {\bibfnamefont {P.}~\bibnamefont
  {Holme}},\ }\href@noop {} {\bibfield  {journal} {\bibinfo  {journal} {EPJ B}\ }\textbf {\bibinfo {volume} {88}},\ \bibinfo
  {pages} {1} (\bibinfo {year} {2015})}\BibitemShut {NoStop}%
\bibitem [{\citenamefont {Pan}\ and\ \citenamefont
  {Saram{\"a}ki}(2011)}]{pan2011path}%
  \BibitemOpen
  \bibfield  {author} {\bibinfo {author} {\bibfnamefont {R.~K.}\ \bibnamefont
  {Pan}}\ and\ \bibinfo {author} {\bibfnamefont {J.}~\bibnamefont
  {Saram{\"a}ki}},\ }\href@noop {} {\bibfield  {journal} {\bibinfo  {journal}
  {Phys. Rev. E}\ }\textbf {\bibinfo {volume} {84}},\ \bibinfo {pages} {016105}
  (\bibinfo {year} {2011})}\BibitemShut {NoStop}%
\bibitem [{\citenamefont {Mattern}(1989)}]{mattern1989virtual}%
  \BibitemOpen
  \bibfield  {author} {\bibinfo {author} {\bibfnamefont {F.}~\bibnamefont
  {Mattern}},\ }in\ \href@noop {} {\emph {\bibinfo {booktitle} {Proceedings of the International Workshop on Parallel and Distributed Algorithms,}}} edited by M. Corsnard {\it et al.}\ (\bibinfo {year} {Elsevier Science Publishers, B.V., North-Holland, 1988})\ pp.\ \bibinfo {pages}
  {215--226}\BibitemShut {NoStop}%
\bibitem [{\citenamefont {Fidge}(1987)}]{fidge1987timestamps}%
  \BibitemOpen
  \bibfield  {author} {\bibinfo {author} {\bibfnamefont {C.~J.}\ \bibnamefont
  {Fidge}},\ }\href@noop {} {\emph {\bibinfo {title} {Timestamps in
  message-passing systems that preserve the partial ordering}}}\ (\bibinfo
  {publisher} {Australian National University. Department of Computer
  Science},\ \bibinfo {year} {1987})\BibitemShut {NoStop}%
\bibitem [{\citenamefont {Mueller}\ and\ \citenamefont
  {Chatterji}(2002)}]{mueller2002analysis}%
  \BibitemOpen
  \bibfield  {author} {\bibinfo {author} {\bibfnamefont {E.~R.}\ \bibnamefont
  {Mueller}}\ and\ \bibinfo {author} {\bibfnamefont {G.~B.}\ \bibnamefont
  {Chatterji}},\ }in\ \href@noop {} {\emph {\bibinfo {booktitle} {Proceedings of the
AIAA?½fs Aircraft Technology, Integration, and Operations (ATIO)
2002 Technical Forum}}}\ (\bibinfo {year}
  {American Institute of Aeronautics and
Astronautics, Los Angeles, 2002}), No. 5866\BibitemShut {NoStop}%
\bibitem [{\citenamefont {Jetzki}(2009)}]{jetzki2009propagation}%
  \BibitemOpen
  \bibfield  {author} {\bibinfo {author} {\bibfnamefont {M.}~\bibnamefont
  {Jetzki}},\ }\href@noop {} {\bibfield  {journal} {\bibinfo  {journal} {PhD
  diss., Aachen University}\ } (\bibinfo {year} {2009})}\BibitemShut {NoStop}%
\bibitem [{\citenamefont {Fleurquin}\ \emph {et~al.}(2013)\citenamefont
  {Fleurquin}, \citenamefont {Ramasco},\ and\ \citenamefont
  {Eguiluz}}]{fleurquin2013systemic}%
  \BibitemOpen
  \bibfield  {author} {\bibinfo {author} {\bibfnamefont {P.}~\bibnamefont
  {Fleurquin}}, \bibinfo {author} {\bibfnamefont {J.~J.}\ \bibnamefont
  {Ramasco}}, \ and\ \bibinfo {author} {\bibfnamefont {V.~M.}\ \bibnamefont
  {Eguiluz}},\ }\href@noop {} {\bibfield  {journal} {\bibinfo  {journal} {Sci.
  Rep.}\ }\textbf {\bibinfo {volume} {3}} (\bibinfo {year} {2013})}\BibitemShut
  {NoStop}%
\end{thebibliography}
\end{document}